\documentclass[preprint]{aastex}

\newcommand{\Msol}{$M_{\odot}$}

\slugcomment{Accepted for publication in ApJ}

\shorttitle{Dynamical Masses of the Binary Brown Dwarf GJ\,569\,Bab}
\shortauthors{Zapatero Osorio et al.}

\begin{document}

\title{Dynamical Masses of the Binary Brown Dwarf GJ\,569\,Bab}

\author{M.\,R$.$ Zapatero Osorio}
\affil{LAEFF-INTA, P.\,O$.$ 50727, E-28080 Madrid, Spain}
\email{mosorio@laeff.esa.es}

\author{B.\,L$.$ Lane}
\affil{MIT Center for Space Research, 70 Vassar Street, Cambridge, MA 02139, USA}
\email{blane@MIT.EDU}

\author{Ya$.$ Pavlenko}
\affil{Main Astronomical Observatory of Ukrainian Academy of Sciences, 
       Zabolotnoho 27, Kyiv-127, 03680 Ukraine}
\email{yp@mao.kiev.ua}

\author{E.\,L$.$ Mart\'\i n}
\affil{Instituto de Astrof\'\i sica de Canarias, E-38200 La Laguna, 
       Tenerife, Spain}
\email{ege@iac.es}

\author{M$.$ Britton}
\affil{Caltech Optical Observatories, Caltech, 
       Pasadena CA 91125, USA}
\email{mbritton@gps.caltech.edu}

\and

\author{S.\,R$.$ Kulkarni}
\affil{Department of Geological \& Planetary Sciences, Caltech, 
       Pasadena CA 91125, USA}
\email{srk@gps.caltech.edu}


%\altaffiltext{2}{Society of Fellows, Harvard University.}

%%%%%%%%%%%%%%%%%%%%%%%%%%%%%%%%%%%%%%%%%%%%%%%%%%%%%%%%%%%%%%%%%%

\begin{abstract}

We have obtained new images and high-resolution ($R$\,$\sim$\,22400) near-infrared (1.2400--1.2575\,$\micron$) spectra of each component of the brown dwarf binary GJ\,569\,Bab using the Adaptive Optics facility of the Keck\,II telescope and the NIRSPEC spectrometer. These data have allowed us to improve the determination of the astrometric orbit and to measure radial velocities of the components. We have used the astrometric and spectroscopic measurements to derive the dynamical mass of each brown dwarf and the systemic velocity of the pair by means of a $\chi^2$ fitting technique. From various considerations, the mass of each component is likely in the range 0.034--0.070\Msol~(GJ\,569\,Bb) and 0.055--0.087\Msol~(GJ\,569\,Ba). This implies that the mass ratio, $q$, of the binary is greater than 0.4, being the most likely value $q$\,=\,0.75--0.85. Adopting 0.072\Msol~as the most conservative location of the substellar limit for solar metallicity, our analysis confirms GJ\,569\,Bb as the {\sl first genuine brown dwarf known without any theoretical assumption}. We have compared the dynamical masses of GJ\,569\,Ba and Bb, and their effective temperatures and luminosities, to the predictions of state-of-the-art theoretical evolutionary isochrones, finding that models exhibit good performance in the regime of high substellar masses if the binary is about a few hundred million years old. However, the surface gravities of GJ\,569\,Ba (M8.5V) and Bb (M9V) derived from our spectral analysis (the observed data have been compared to the latest synthetic spectra) appear to be smaller than the values provided by the evolutionary models.

\end{abstract}


%% subject header.  Objects should be in the appropriate "individual"
%% headers (e.g. quasars: individual, stars: individual, etc.) with the
%% additional provision that the total number of headers, including each
%% individual object, not exceed six.  The \objectname{} macro, and its
%% alias \object{}, is used to mark each object.  The macro takes the object
%% name as its primary argument.  This name will appear in the paper
%% and serve as the link's anchor in the electronic edition if the name
%% is recognized by the data centers.  The macro also takes an optional
%% argument in parentheses in cases where the data center identification
%% differs from what is to be printed in the paper.

\keywords{stars: low-mass, brown dwarfs ---
          stars: individual (GJ\,569\,B) ---
          stars: fundamental parameters ---
          binaries: visual
}

\section{Introduction}

Since the discovery of the first brown dwarfs (Nakajima et al$.$ \cite{nakajima95}; Rebolo, Zapatero Osorio, \& Mart\'\i n \cite{rebolo95}), with masses below the hydrogen burning mass limit (Kumar \cite{kumar63}; Grossman, Hays, \& Graboske \cite{grossman74}; Burrows et al.~\cite{burrows93}; D'Antona \& Mazzitelli \cite{dantona94}), many efforts have been devoted to study their physical properties and formation processes. The frequency and characteristics of binary and multiple brown dwarf systems and brown dwarf--star pairs may provide key information on stellar/substellar/planetary formation. With the advent of new ground-based observational techniques, like the Adaptive Optics and coronography, and the high spatial resolution imaging capability of the Hubble Space Telescope, it has been possible to resolve very low mass companions around stars and binary brown dwarfs (e.g., Mart\'\i n, Brandner, \& Basri \cite{martin99}; Close et al$.$ \cite{close02a}, \cite{close02b}; Potter et al$.$ \cite{potter02}; Burgasser et al$.$ \cite{burgasser03}; Gizis et al$.$ \cite{gizis03}; McCaughrean et al$.$ \cite{mccaughrean04}). One spectroscopic binary brown dwarf has also been discovered in the young Pleiades open cluster (Basri \& Mart\'\i n \cite{basri99}). All these surveys indicate that substellar binaries are not rare, and that brown dwarfs are commonly found in orbit around low mass stars. These are rather nearby systems; the visual binary brown dwarfs known so far typically present projected separations between 0.9 and $\sim$10\,AU (orbital periods of 3--30\,yr or larger), permitting orbital mapping over the coming decades. 

Eclipsing binaries have often served as valuable tools for the validation of structure and evolutionary models (e.g., Delfosse et al$.$ \cite{delfosse00}). Accurate substellar mass, radius and luminosity measurements provide a crucial test of our understanding of substellar physics. To the best of our knowledge, no eclipsing brown dwarfs are identified so far. In the absence of eclipsing substellar systems, nearby visual brown dwarf pairs become benchmark objects to evaluate theoretical models in the substellar mass regime, because astrometric and spectroscopic studies will yield highly accurate dynamical masses of the components, although the majority of these binaries tends to have long orbital periods. The binary GJ\,569\,Bab is one of the very low mass pairs with the shortest periods known to date (2.4\,yr, Lane et al$.$ \cite{lane01a}). It can be used to calibrate state-of-the-art evolutionary isochrones.

GJ\,569\,Bab is a binary dwarf moving around the chromospherically active M2.5V-type star GJ\,569\,A (Hipparcos distance of 9.8\,pc, Perryman et al$.$ \cite{perryman97}). The presence of a ``single'', very red, proper motion companion at a separation of 5\arcsec~from the primary star was first reported by Forrest, Skrutskie, \& Shure \cite{forrest88}. Follow-up low-resolution spectroscopy obtained by Henry \& Kirkpatrick \cite{henry90} confirmed its cool nature (it was classified as an M8.5V dwarf). Mart\'\i n et al$.$ \cite{martin00} resolved the companion into two separate objects (GJ\,569\,Ba and GJ\,569\,Bb) using Adaptive Optics (AO) observations with the Keck\,II telescope. More recently, we obtained AO images and low-resolution spectra of each of the components (Lane et al$.$ \cite{lane01a}, hereafter Paper~I), which have allowed us to determine the astrometric orbital parameters of the binary (see Table~\ref{astrometry}) and the spectral types of GJ\,569\,Ba (M8.5V) and GJ\,569\,Bb (M9V). Similar astrometric results obtained from speckle interferometry were also reported by Kenworthy et al$.$ \cite{kenworthy01}. In this paper, we present high-resolution spectra of the resolved components obtained at different orbital epochs and the derivation of the dynamical masses of GJ\,569\,Ba and Bb. For the first time, one brown dwarf is unambiguosly confirmed without any theoretical model. This has allowed us to carry out a critical evaluation of substellar evolutionary model predictions.

\section{Observations and data reduction}

\subsection{Adaptive Optics imaging and improved astrometric orbit \label{improvedorbit}}

In addition to the astrometry presented in Paper~I, we have observed the binary GJ\,569\,Bab on two more occasions on 2001 Jun 28 and Sep 01. We used the 10-m Keck II AO facility (Wizinowich et al$.$ \cite{wizinowich88}), and the slit-viewing camera (SCAM) associated with the NIRSPEC instrument (McLean et al$.$ \cite{mclean98}). The pixel scale of the 256\,$\times$\,256 Rockwell HgCdTe array of SCAM was 0\farcs0168. Exposures were taken in the $K'$-band. The observational strategy, data reduction, and the analysis of the astrometry are all described in Paper~I. We list in Table~\ref{astrometry} the most recent relative positions of GJ\,569\,Bb with respect GJ\,569\,Ba as a function of the Modified Julian Date (MJD). 

\begin{figure}
\epsscale{1.0}
\plotone{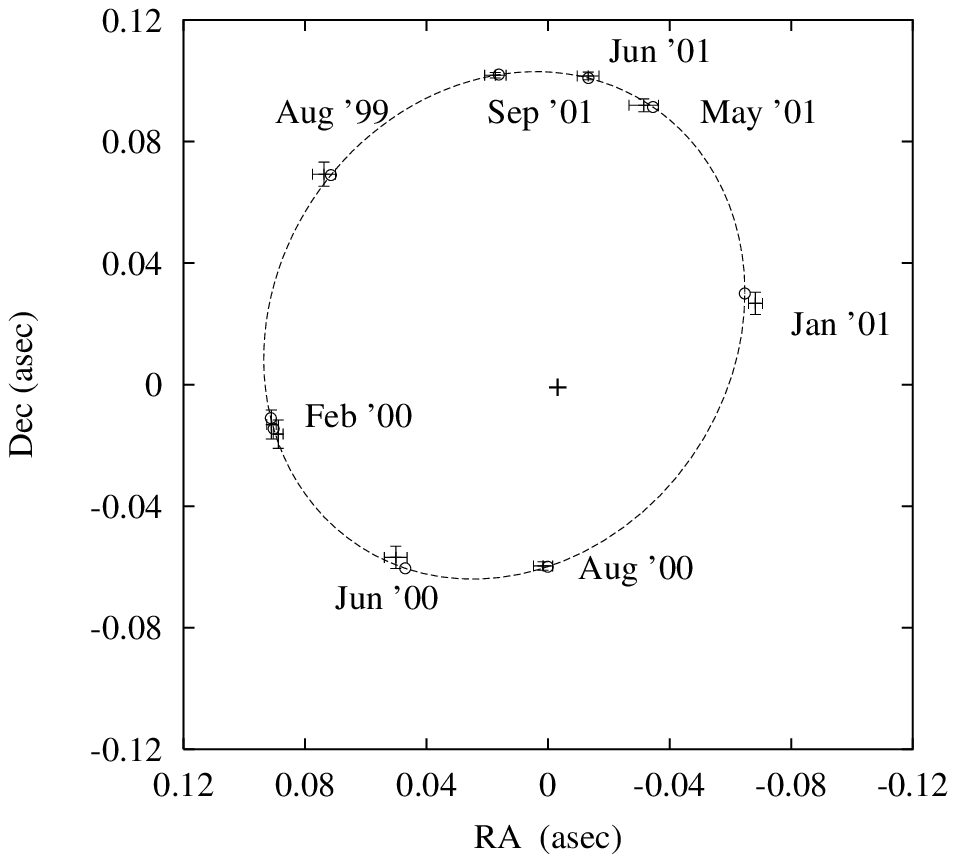}
\caption{The new astrometric data (2001 Jun and Sep) are plotted together with the best-fit orbit (dashed ellipse) and the previous astrometry (Lane et al$.$ \cite{lane01a}). Error-bar crosses denote measurements and circles indicate the predicted location on the orbit at the time of the observations.  North is up, and east is to the left.\label{orbiteps}}
\end{figure}

These new astrometric data along with the previous astrometry allow us to determine the orbit of GJ\,569\,Bab to a certain degree of accuracy by fitting a Keplerian model. A description of the procedure by which the best Keplerian solution is obtained is fully detailed in Paper~I. The updated best-fit parameters and their 1$\sigma$ uncertainties are given in Table~\ref{orbit} (columns 2 and~3), where we also provide the previous measurements for comparison (columns 4 and~5). The agreement is excellent, and the new and ``old'' values are very similar within the error bars. We note that the uncertainties of the updated parameters are reduced as compared to Paper~I. This suggests that the relative orbit of the GJ\,569\,Bab pair is now determined to a high degree of confidence. The visual fit orbit is shown together with all the available astrometry in Fig.~\ref{orbiteps}. 

It is important to note that the astrometric solution of the orbit in Paper~I has an ambiguity of 180$^{\circ}$ in the determination of the position angle of the ascending node, $\Omega$. This parameter sets how the plane of the orbit points in or out of the plane of the sky. From purely positional measurements we are able to identify the two nodes of the orbit, but it is not known which is the ascending node. Because the argument of periapsis, $\omega$, is measured from the position of the ascending node, it also shares a similar ambiguity. This uncertainty does not affect the determination of the other orbital parameters or the total mass of the binary, and is only removed with spectroscopic data. The radial velocities that we have measured for the GJ\,569\,Bab system (Section~\ref{velocities}) confirm that the angle of the node published in Paper~I corresponds to the position of the descending node. In column 2 of Table~\ref{orbit} we provide the correct mean values for $\Omega$ and $\omega$. 

From the updated astrometric orbit of the pair, the best-fit values of the period and semi-major axis correspond to a total mass of 0.125\Msol~with a 3$\sigma$ uncertainty of $\pm$0.015\Msol. This is a purely Keplerian dynamical result, and no evolutionary models are involved in the calculations. The 3$\sigma$ upper mass limit of GJ\,569\,Bab is 0.140\Msol. As discussed in Paper~I, the similarity of the photometric and spectroscopic properties between GJ\,569\,Ba and Bb suggests that the mass ratio, $q$, of the pair is close to equal. This sets the 3$\sigma$ upper mass limit of the smallest object to be 0.070\Msol~(0.065\Msol~at 1$\sigma$), and the lower mass limit of the largest object to be 0.055\Msol~(0.060\Msol~at 1$\sigma$). Magazz\`u, Mart\'\i n \& Rebolo~\cite{magazzu93} did not detect lithium in the optical spectrum of GJ\,569\,Bab. Their data are dominated by the more massive component, which is brighter. Theoretical models (e.g., Burrows et al.~\cite{burrows97}; Baraffe et al.~\cite{baraffe98}) predict that objects more massive than 0.055\Msol~have efficiently depleted their atmospheric lithium at ages older than a few hundred million years. In this sense, models and our observations appear to be in agreement. We will discuss lithium in each component of the binary in a forthcoming paper. To investigate further the reliability of the models, we must know the individual masses of GJ\,569\,Ba and Bb. The radial velocity measurements of Section~\ref{velocities} shall constrain better the mass of each component, providing a better estimate of $q$.

\subsection{Adaptive Optics high-resolution spectroscopy}

We have obtained $J$-band spectra with the Keck II telescope, the AO correcting system, and NIRSPEC, a cross-dispersed, cryogenic echelle spectrometer employing a 1024\,$\times$\,1024 ALADDIN InSb array detector. These observations were carried out on four different occasions within the same orbital period during 2001; three of them coincide with astrometric observations. In the echelle mode, we selected the NIRSPEC-3 ($J$) filter, and an entrance slit width of 0\farcs43 (i.e, 3 pixels along the dispersion direction of the detector). The length of the slit was 12\arcsec. For the present study we used only the spectral order that covers the following wavelength range: 1.2400--1.2575\,$\mu$m. This instrumental setup provided a nominal dispersion of 0.179\,\AA\,pix$^{-1}$ (4.3\,km\,pix$^{-1}$), and a final resolution of 0.55\,\AA, which corresponds to a resolving power $R$\,$\sim$\,22500 at 1.2485\,$\mu$m. The journal of observations is given in Table~\ref{spectroscopy} and shows the observing dates, exposure times and the air mass range of the observations. 

The slit was aligned with GJ\,569\,Ba and GJ\,569\,Bb so that the two objects were observed simultaneously. Spectra were collected at two different positions along the entrance slit. In order to remove atmospheric telluric absorptions, the near-infrared featureless stars HR\,5567 (A0V), HR\,5931 (A0V) and SAO\,101508 (F8III) were observed very close in time and in air mass (within 0.1 air masses). White-light spectra obtained with the same instrumental configuration were used for flat-fielding the target data. The dispersion solutions for the May, June and September data were derived from the arc lamp lines of Ar, Kr, and Xe, which were systematically taken after observing GJ\,569. No calibrating arc lamps were observed in August. 

\begin{figure}
\epsscale{1.0}
\plotone{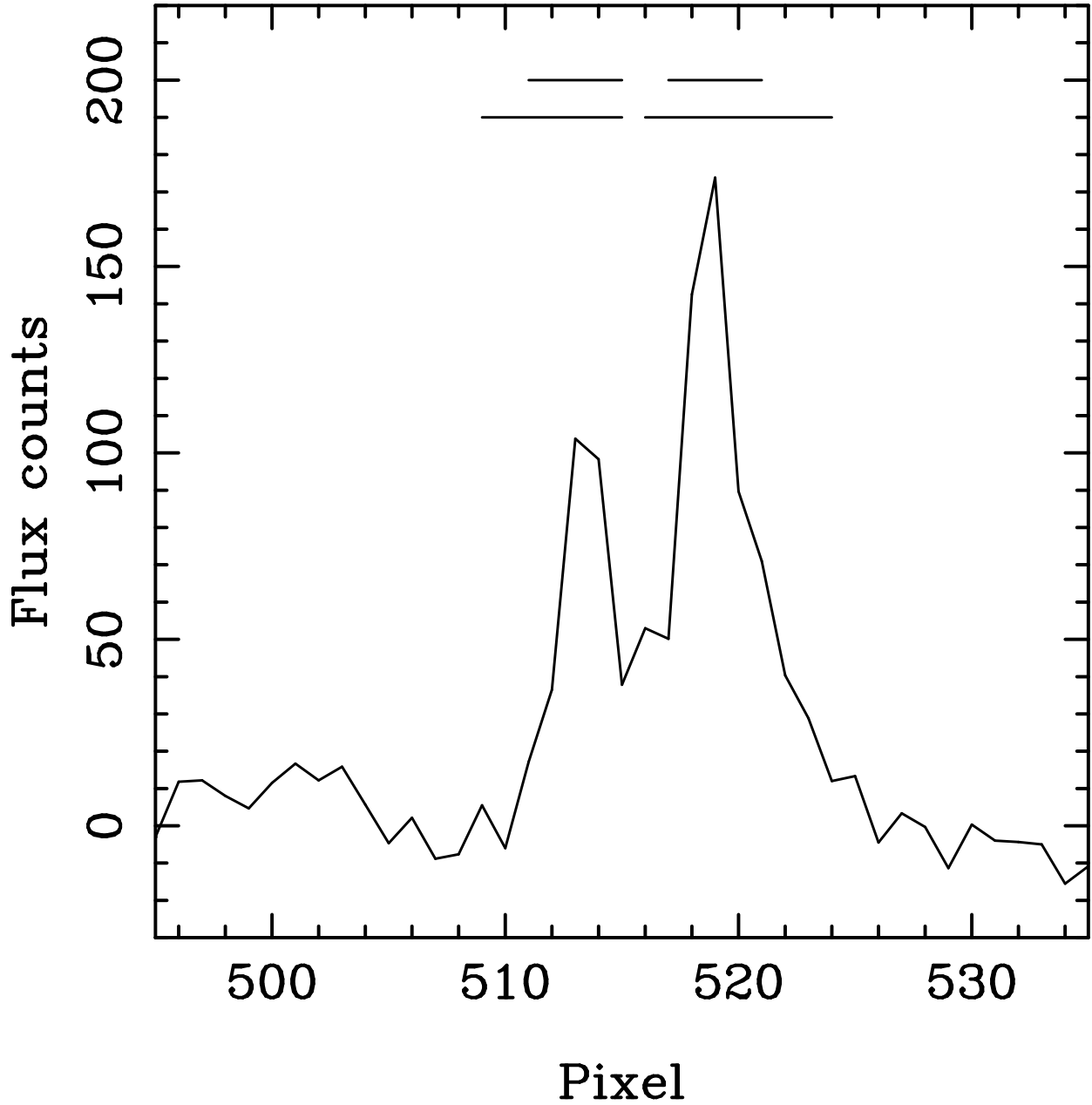}
\caption{Cut along the spatial axis of the 2001 May two-dimension spectra of GJ\,569\,Ba (brightest peak) and Bb. A total of 6 columns are averaged. The smallest and largest apertures used to extract the individual spectra are shown to the top. \label{apertures}}
\end{figure}

Raw spectra were reduced following conventional techniques in the near-infrared and using the REDSPEC software, which was particularly designed at the University of California for the analysis of NIRSPEC data. We also reduced and extracted spectra using the packages TWODSPEC and ONEDSPEC within the IRAF\footnote{IRAF is distributed by National Optical Astronomy Observatory, which is operated by the Association of Universities for Research in Astronomy, Inc., under contract with the National Science Foundation.} environment, obtaining very similar results. Nodded images were subtracted to remove sky background and dark current. We selected different apertures to extract the individual spectra (between 4 and 6 pixels for GJ\,569\,Bb, and between 4 and 8 pixels for GJ\,569\,Ba). Figure~\ref{apertures} shows the cut along the spatial direction of one of our two-dimension spectra. As can be seen from the location and distribution of the apertures in the Figure, the cross-contamination between the extracted data of the two components is less than $\sim$10\%~(except for the September data). We have checked that the relative radial velocity measurements obtained for the various apertures are consistent with each other within the error bars (see discussion of Sect$.$ \ref{velocities}). Larger extraction apertures are useless for radial velocity measurements because the spectrum of each component of the binary is significantly contaminated by the presence of the companion. Extracted spectra were divided by normalized flat-fields, and calibrated in wavelength. The 1$\sigma$ dispersion of the fourth-order polynomial fit was 0.02\,\AA~(3.6\%~of the final resolution). We also removed some fringing that is internal to NIRSPEC and appears sometimes when using this instrument behind the AO system. The spectra of hot stars were used for division into the corresponding science spectra. These stars do not show any intrinsic atmospheric feature in the spectral coverage of the echelle order at 1.2485\,$\mu$m. Finally, we multiplied the science spectra by the black body spectrum for the temperatures of 9480\,K and 11200\,K, which corresponds to the A0V and F8III classes, respectively (Allen \cite{allen00}).

\begin{figure*}
\epsscale{1.0}
\plottwo{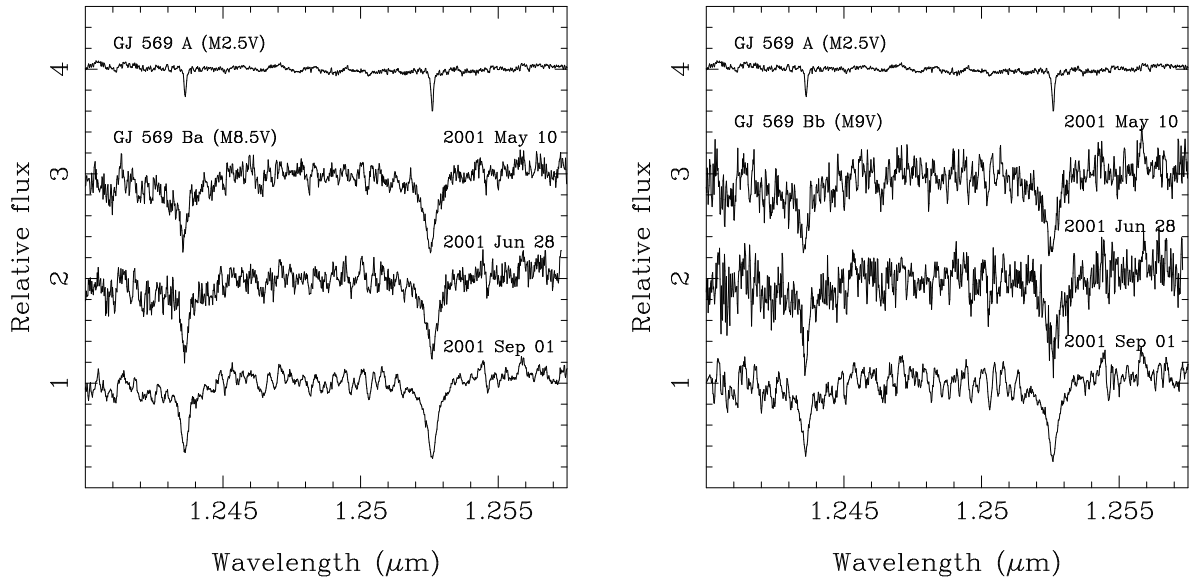}{f2b.ps}
\caption{NIRSPEC spectra of the GJ\,569 triple system. The spectrum of the primary star GJ\,569\,A was obtained in 2001 Jun 28. The strong absorption features are due to the atomic K\,{\sc i} doublet at 1.2436 and 1.2526\,$\mu$m. Data are displaced upwards by one for clarity. \label{spectra}}
\end{figure*}

The final spectra are shown in Fig.~\ref{spectra}. No heliocentric or radial velocity corrections have been applied. The August data are not displayed because they lack a reliable wavelength calibration. The most noticeable spectral features are due to the absorption K\,{\sc i} resonance doublet at 1.2436 and 1.2526\,$\mu$m. The spectrum of the early M-type star GJ\,569\,A shows narrow atomic lines and no other relevant features are apparent. However, the K\,{\sc i} lines of the late-M dwarfs GJ\,569\,Ba and C appear quite broad, as expected for rather cool atmospheres. We note that these atomic lines are moderately stronger in GJ\,569\,Bb than in Ba, which is indicative of the slightly cool nature of the smallest object. Many other absorption features are likely due to H$_2$O, TiO, FeH and CrH, as discussed in Sect$.$~\ref{synthetic}. 

There are clear advantages in determining radial velocities from the spectra of Fig.~\ref{spectra}. On the one hand, this wavelength interval is almost completely free of strong Earth's telluric absorption lines that might introduce uncertainties in the Doppler shift calculation. There are a few of these telluric lines in the red part of the spectra. Nevertheless, we are confident that the science spectra have good cancellation of these terrestrial atmospheric features. In addition, the wavelength solution of the spectral order at 1.2485\,$\mu$m is reliable because it is based on seven arc lamp lines distributed all across the dispersion axis of the detector. Finally, the science spectra have relatively good signal-to-noise ratio at these wavelengths, and display numerous characteristic features that may yield accurate radial velocity measurements.

\section{Analysis}

\subsection{Radial velocities \label{velocities}}

The Doppler shifts of GJ\,569\,Bb were determined by a Fourier cross-correlation technique, using the spectrum of GJ\,569\,Ba as a template. The relative velocities of GJ\,569\,Bb cross-correlated against GJ\,569\,Ba obtained for the four observing nights are given in the second column of Table~\ref{vrad}. According to the measurements, GJ\,569\,Bb is moving toward us, while GJ\,569\,Ba is moving further away at the time of the observations. This has allowed us to determine unambiguosly the position of the ascending node angle of the orbit of the pair (Table~\ref{orbit}). We have measured the observed radial velocities of each component from the centroids of the K\,{\sc i} lines, obtained by marking two continuum points around the lines and fitting the profiles within {\sc iraf}. These velocities were corrected for the rotation of the Earth, the motion of the Earth about the Earth-Moon barycenter, and the orbit of the barycenter about the Sun to yield heliocentric radial velocities, which are listed in columns 3 and 4 of Table~\ref{vrad}. Heliocentric velocities cannot be derived from the August data because the spectra lack a reliable wavelength calibration.  

The precision of all the velocities may be judged by the standard deviation of the measurements for different regions of the spectra. The cross-correlation technique was able to achieve precisions of about 1/4 pixel (an accuracy of the order of $\pm$1.1\,km\,s$^{-1}$) for good signal-to-noise data. The velocity accuracy became a factor of 1.5 to 2 worse when analizing the data of GJ\,569\,Bb, because this object is fainter than GJ\,569\,Ba and its spectra show lower signal-to-noise ratios (Fig.~\ref{spectra}). Similar uncertainties are affecting the radial velocities derived from the centroids of the atomic lines. We have measured the heliocentric velocity of the star GJ\,569\,A using its 2001 June spectrum, and have found it to be $-$8.0\,$\pm$1.1\,km\,s$^{-1}$. Various authors have investigated the velocity variability of GJ\,569\,A during the last two decades, concluding that this is a rather stable star over a period of ten years with a mean heliocentric velocity of $-$7.17\,$\pm$0.28\,km\,s$^{-1}$ (Marcy, Lindsay \& Wilson \cite{marcy87}; Marcy \& Benitz \cite{marcy89}), $-$8.6\,$\pm$1.0\,km\,s$^{-1}$ (Barbier-Brossat \& Figon \cite{barbier00}), and $-$7.2\,$\pm$0.4\,km\,s$^{-1}$ (Nidever et al$.$ \cite{nidever02}). Our measurement and its error bar are in agreement with the values seen in the literature. We are confident that the velocities of Table~\ref{vrad} are accurate to the claimed uncertainties. 

\subsection{Atmospheric gravity and rotation from spectral synthesis \label{synthetic}}

By comparing our observed spectra to synthetic energy distributions, we can obtain information on the atmospheric temperatures, gravities and rotation of GJ\,569\,Ba and Bb. Theoretical spectra were computed in a classical framework, i.e., assuming LTE, plane-parallel media, and no sinks and sources of energy in the atmosphere. We used the WITA6 code (Pavlenko \cite{pav00}), which takes into account a total of 100 species. The equations of ionisation-dissociation equilibrium were solved for a plasma made of neutral atoms, ions and molecules. The constants of the formulas for chemical balance were taken from Tsuji \cite{tsuji73} and Gurvitz et al$.$ \cite{gurvitz82}. Line lists come from different sources, depending on the species. Atomic line lists were taken from VALD (Kupka et al$.$ \cite{kupka99}), lines of TiO from Plez \cite{plez98}, CN from Kurucz \cite{kurucz93}, CrH and FeH from Burrows et al$.$ \cite{burrows02} and Dulick et al$.$ \cite{dulick03}, respectively, and H$_2$O from Partrige \& Schwenke \cite{partrige97}. The partition function of H$_2$O was computed as in Pavlenko \cite{pav02} to properly account for the splitting of the lines. We adopted the Voigt profile for the shape of each molecular and atomic line; the damping constants were computed using data from the original databases, and if no values were published, computed following an approximation of Uns\"old \cite{unsold55}. Due to the low temperatures of the atmospheres of late-M dwarfs, Stark broadening may be discarded, whereas pressure broadening prevails.

\begin{figure}
\epsscale{1.0}
\plotone{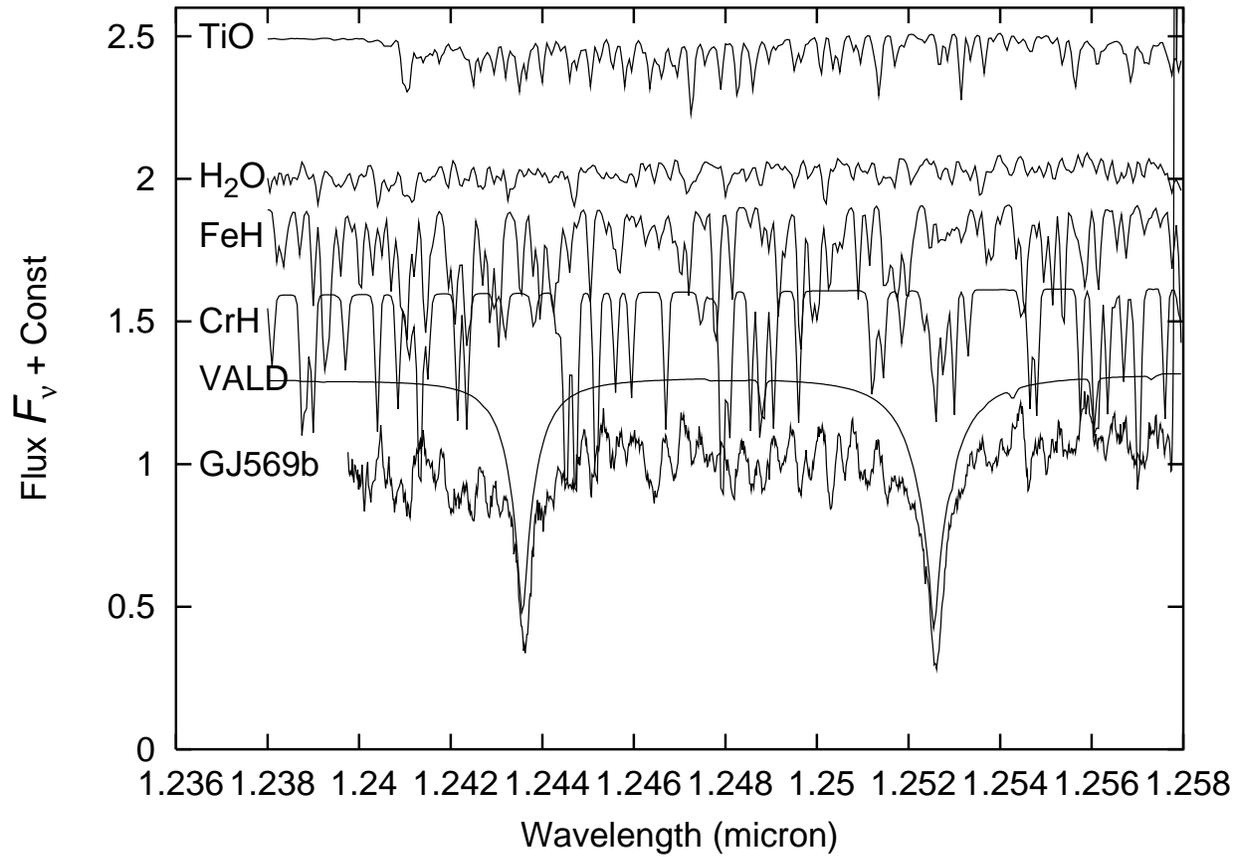}
\caption{Contribution of different molecular and atomic species to the opacity in the spectral region of our observations. \label{species}}
\end{figure}

A grid of 20 synthetic spectra was computed for solar metallicity, {\sc dusty} model atmospheres of Allard et al$.$~\cite{allard01}, effective temperatures, $T_{\rm eff}$, of 2600, 2400, 2300, 2000\,K, and gravities log\,$g$\,=\,4.0, 4.5, 5.0, 5.5, 6.0 (cm\,s$^{-2}$). We have also computed grids for the {\sc cond} model atmospheres of Allard et al$.$~\cite{allard01}, and the very recent {\sc ucm} models of Tsuji \cite{tsuji02} for the critical temperature of 1800\,K. The three sets of models differ in the treatment of dust (see the references above). Late-M and early-L dwarfs are supposed to have dusty upper atmospheres, with dust particles affecting the thermal structure. Hence, the {\sc cond} models, which assume a complete dust-segregation, are not a good approximation for the case of GJ\,569\,Bab. Theoretical spectra were obtained in the wavelength range of our observations, with wavelength step of 0.5\,\AA, and with microturbulent velocity of 2\,km\,s$^{-1}$. Element abundances are as in Anders \& Grevesse \cite{anders89}. The contributions of various molecular and atomic species to the total opacity in the 1.2400--1.2575\,$\micron$ spectral region are shown in Fig$.$~\ref{species}, from which we see the absorption due to K\,{\sc i} dominates. The remaining features of the observed spectra are likely due to H$_2$O, CrH and FeH. However, the absorption at 1.2462\,$\micron$ does not appear to have a counterpart in any of the molecular data of the figure. 

\begin{figure*}
\epsscale{1.0}
\plottwo{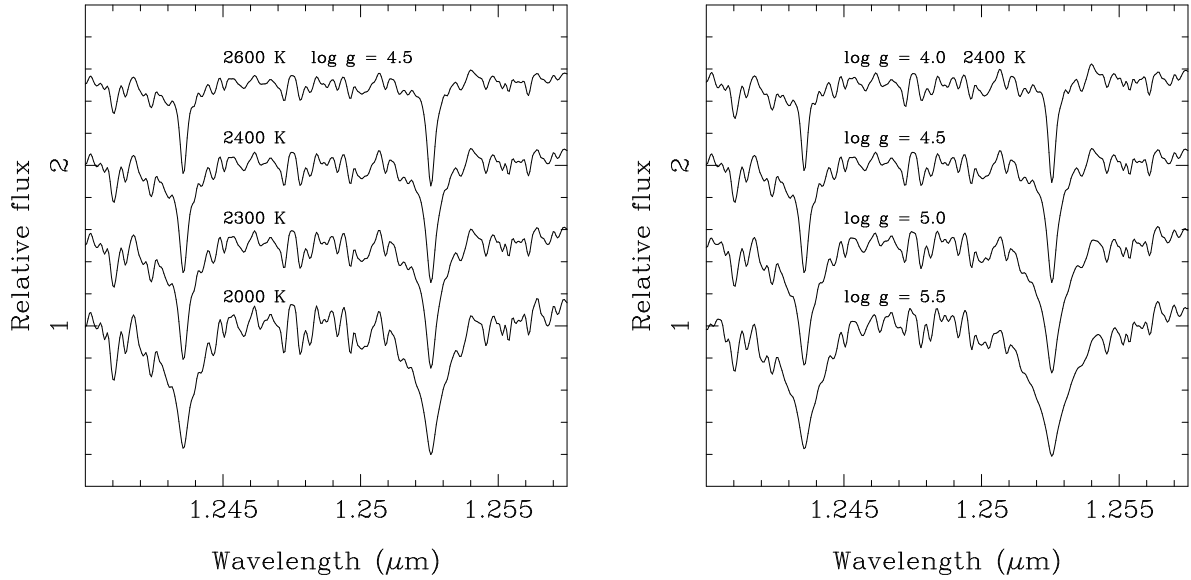}{f4b.ps}
\caption{Dependence of theoretical spectra on $T_{\rm eff}$ {\sl (left panel)} and log\,$g$ {\sl (right panel)}. CrH is not included in these computations, and van der Waals broadening due to collisions with H and H$_2$ is considered. Spectra, computed using the {\sc dusty} model atmospheres of Allard et al$.$ \cite{allard01}, have been rebinned to the resolution of our data and rotationally broadened by 30\,km\,s$^{-1}$. An offset of 0.5 units is added to each spectrum for clarity.  \label{theory}}
\end{figure*}

The effect of changing $T_{\rm eff}$ and log\,$g$ in our computations is shown in Fig$.$~\ref{theory}. We note that CrH was excluded from our final synthetic spectra, because it appears too strong in the spectral region of the figure (see Fig$.$~\ref{species}). Our spectroscopic data are better reproduced by computations that do not include the opacities of this molecule. As seen from Fig$.$~\ref{theory}, temperature has a relatively larger impact on the energy distributions than has gravity. It is also obvious that the K\,{\sc i} lines are remarkably sensitive to both temperature and gravity, i.e., the doublet becomes systematically weaker at high temperatures and low gravities (e.g., see McGovern et al$.$ \cite{mcgovern04}), while the main molecular species at these wavelengths do not appear to change notably with gravity. This justifies the use of the K\,{\sc i} atomic lines to estimate the surface gravity of GJ\,569\,Ba and Bb. 

\begin{figure}
\epsscale{1.0}
\plotone{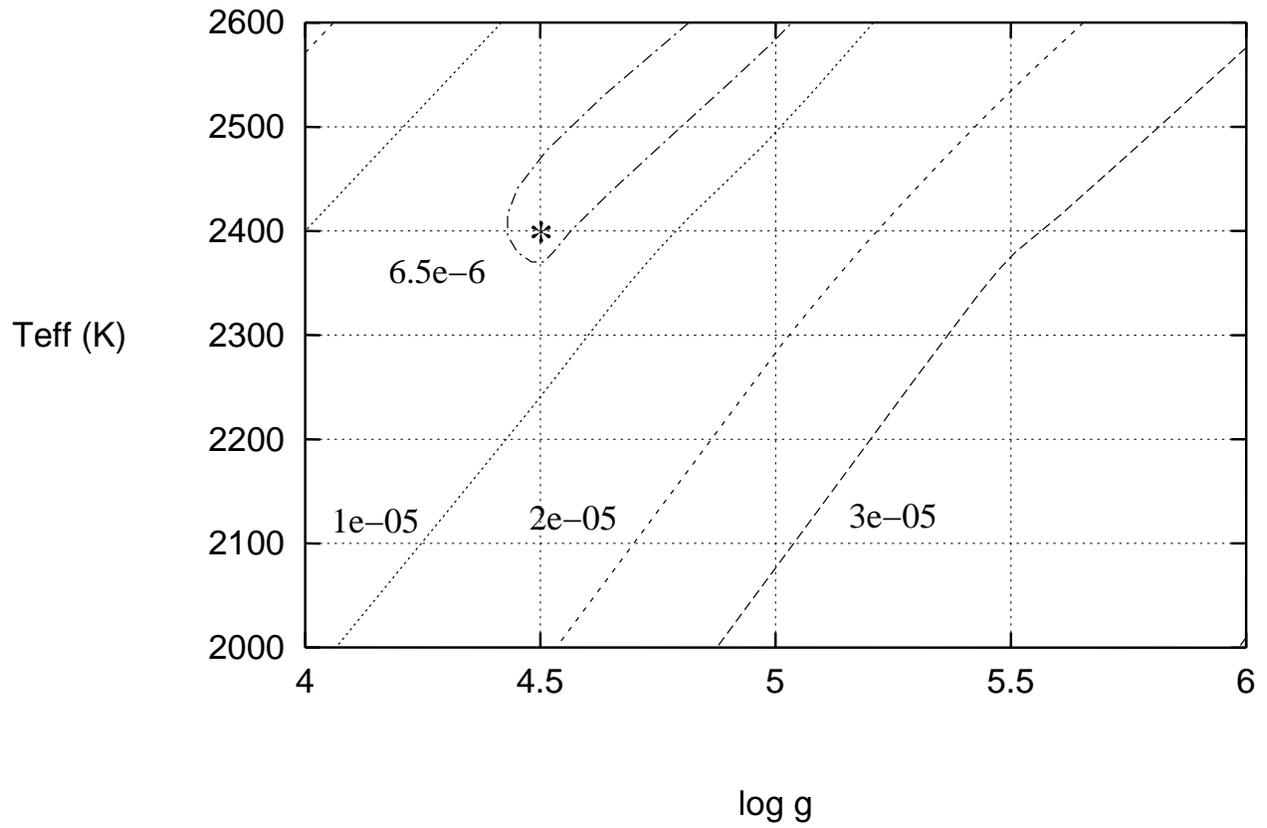}
\caption{The contours of the ``S'' function, defined by Pavlenko et al$.$~\cite{pav04}, against $T_{\rm eff}$ and log\,$g$ (cm\,s$^{-2}$) for the {\sc dusty} model atmospheres. The best-fit parameters (4.5, 2400\,K) are given by the asterisk.  \label{minima}}
\end{figure}

\begin{figure*}
\epsscale{1.0}
\plottwo{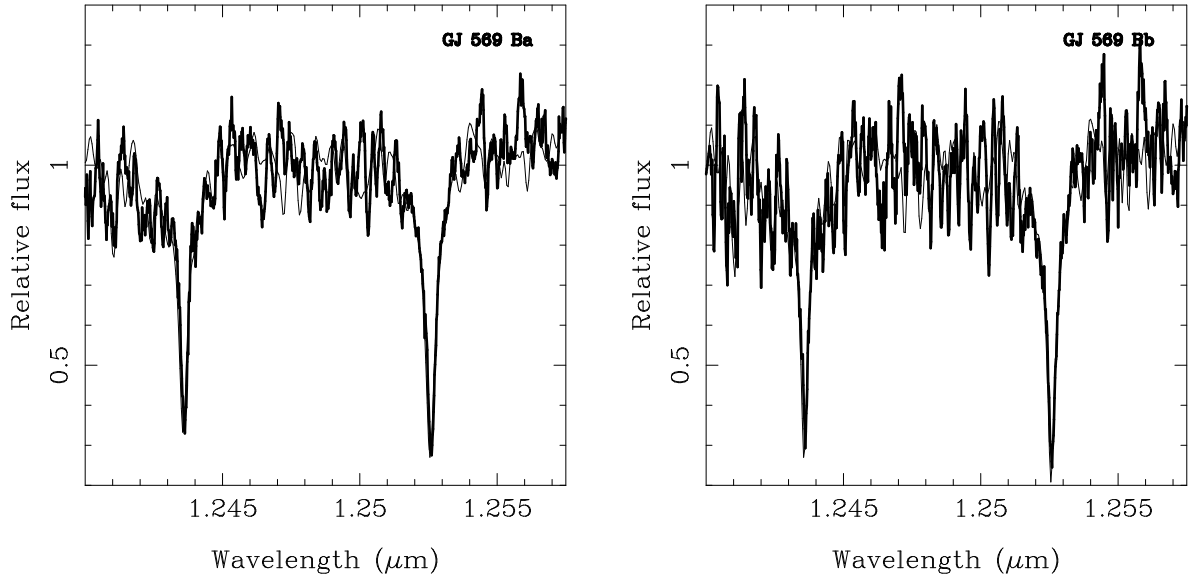}{f6b.ps}
\caption{Spectral data of GJ\,569\,Ba and Bb (thick lines) and the best-fit {\sc dusty} theoretical spectrum (2400\,K, log\,$g$\,=\,4.5, thin lines). Note that while the alkali lines are nicely reproduced by the theoretical data, the molecular absorptions are not well matched by the computations. \label{bestfitspectra}}
\end{figure*}

Our observed spectra have been compared to the various grids of synthetic data, which were computed including the broadening of the alkali lines due to van der Waals collisions with H and H$_2$ (important in cool, neutral atmospheres). The van der Waals treatment is an approximation applicable to the K\,{\sc i} doublet at 1.25\,$\micron$, because these are weak lines (e.g., Burrows, Marley, \& Sharp \cite{burrows00}). The comparison has been performed as in Jones et al$.$ \cite{jones02}, Pavlenko \& Jones \cite{pavjones02} and Pavlenko et al$.$ \cite{pav04}. The analysis procedure described by these authors is based on a minimization technique that provides the best-fit $T_{\rm eff}$, log\,$g$ and $v_{\rm rot}$\,sin\,$i$ (line broadening due to atmospheric rotation), where $i$ is the inclination angle of the rotation axis. Of the three sets of model atmospheres, the {\sc dusty} models yield the smallest values of the minimization function. The best-fit parameters are as follows: for both GJ\,569\,Ba and Bb, $T_{\rm eff}$\,=\,2400\,K and log\,$g$\,=\,4.5 ({\sc dusty} models), and $T_{\rm eff}$\,=\,2500\,K and log\,$g$\,=\,4.5 ({\sc ucm} models). Tsuji's \cite{tsuji02} models provide higher $T_{\rm eff}$ because they predict lower temperatures at the atmospheric layers where lines are produced. A plot depicting the ($T_{\rm eff}$, log\,$g$) contours of the minimization function is shown in Fig$.$~\ref{minima}, and Fig$.$~\ref{bestfitspectra} displays the best-fit synthetical spectra together with the observed data. We note that while the alkali lines are nicely reproduced by the theoretical data, the molecular absorptions are not well matched by the computations. 

Unfortunately, we do not have a reference slow-rotator spectrum (of similar spectral type and observed with the same instrumentation) to measure the rotational velocities of GJ\,569\,Ba and Bb. The fitting technique of Pavlenko et al$.$ \cite{pav04} provides estimates for these velocities. Despite the unknown inclination angles of the two objects, our spectral analysis suggests that GJ\,569\,Ba and Bb are fast rotators, with Ba ($v_{\rm rot}$\,sin\,$i$\,$\sim$\,37\,$\pm$\,15\,km\,s$^{-1}$) possibly rotating slightly faster than Bb ($v_{\rm rot}$\,sin\,$i$\,$\sim$\,30\,$\pm$\,15\,km\,s$^{-1}$). We note that there is a large uncertainty associated to these measurements, mainly due to the poor knowledge of the molecular opacities and the uncertainty in the $T_{\rm eff}$ determination, which is about 100\,K. A moderately cooler temperature than that of GJ\,569\,Ba and a similarly high rotational velocity may explain the K\,{\sc i} lines broadening and the dissimilarities of GJ\,569\,Bb. These rapid velocities are not surprising since many field low-mass stars and brown dwarfs display high values of $v_{\rm rot}$\,sin\,$i$ as indicated by the spectroscopic measurements of Basri \& Marcy \cite{basri95}, Tinney \& Reid \cite{tinney98}, Basri \cite{basri01} and Reid et al$.$ \cite{reid02}. According to evolutionary models (e.g., Baraffe et al$.$ \cite{baraffe03}), the radius expected for 0.060--0.070\Msol~brown dwarfs is in the range 0.130--0.092\,$R_\odot$ for ages between 300 and 1000\,Myr. This would imply short rotation periods between 3\,sin\,$i$ and 5\,sin\,$i$ hours. Some field and cluster brown dwarfs of similar spectral classes have been identified with such rapid rotations, like Kelu\,1 (1.8\,h, Clarke, Tinney, \& Covey \cite{clarke02}), and S\,Ori\,27, S\,Ori\,28 and S\,Ori\,45 of the $\sigma$\,Orionis cluster (2.8, 3.3 and $\sim$3\,h, respectively, Caballero et al$.$ \cite{caballero04}; Zapatero Osorio et al$.$ \cite{osorio03}). 

Because the rotation velocities of GJ\,569\,Ba and Bb appear to be different, and assuming that both objects might have similar inclination rotation angles, it is unlikely that the two components are tidally pseudo-synchronized (i.e., the angular velocity of rotation is similar to the angular velocity of the companion crossing the sky during periastron passage). This contrasts with the case of the Pleiades member PPl\,15 (120\,Myr), which is, to the best of our knowledge, so far the only known spectroscopic binary brown dwarf (Basri \& Mart\'\i n \cite{basri99}). The rotation of each of the components is similar and slow (10\,km\,s$^{-1}$), suggesting that the two objects are tidally pseudo-synchronized regardless their youth. The very short (the semi-major axis is 30 times smaller than that of GJ\,569\,Bab) and eccentric ($e$\,=\,0.42) orbit of PPl\,15 may also contribute to its pseudo-synchronism. 

We have also compared our spectroscopic data of GJ\,569\,Ba and Bb to computations carried out considering line broadening due to collisions with H (collisions with H$_2$ were excluded). This comparison yields higher values of the minimization function, the same $T_{\rm eff}$, similar rotational velocities within 5\,km\,s$^{-1}$, and higher log\,$g$ values (log\,$g$\,=\,5.0), for both the {\sc dusty} and {\sc ucm} models. Hence, we will adopt log\,$g$\,=\,4.5--5.0 as the best surface gravity estimates for GJ\,569\,Ba and Bb. This result is in rough agreement with the recent calculations done by Gorlova et al$.$ \cite{gorlova03}. Concerning $T_{\rm eff}$, the range 2400--2500\,K is indeed very close to the estimates of Paper~I, where we found 2440\,$\pm$\,100\,K for GJ\,569\,Ba and 2305\,$\pm$\,100\,K for GJ\,569\,Bb. We will use these values throughout the present paper.

\section{The mass of GJ\,569\,Ba and GJ\,569\,Bb \label{mass_determination}}
Radial velocities combined with the knowledge of the orbital parameters from astrometry robustly constrain the individual masses of visual binary members. However, in the case of the pair GJ\,569\,Bab there is an additional parameter that has to be considered: its orbital motion around the center of mass of the triple system GJ\,569\,ABab.  

\begin{figure}
\epsscale{1.0}
\plotone{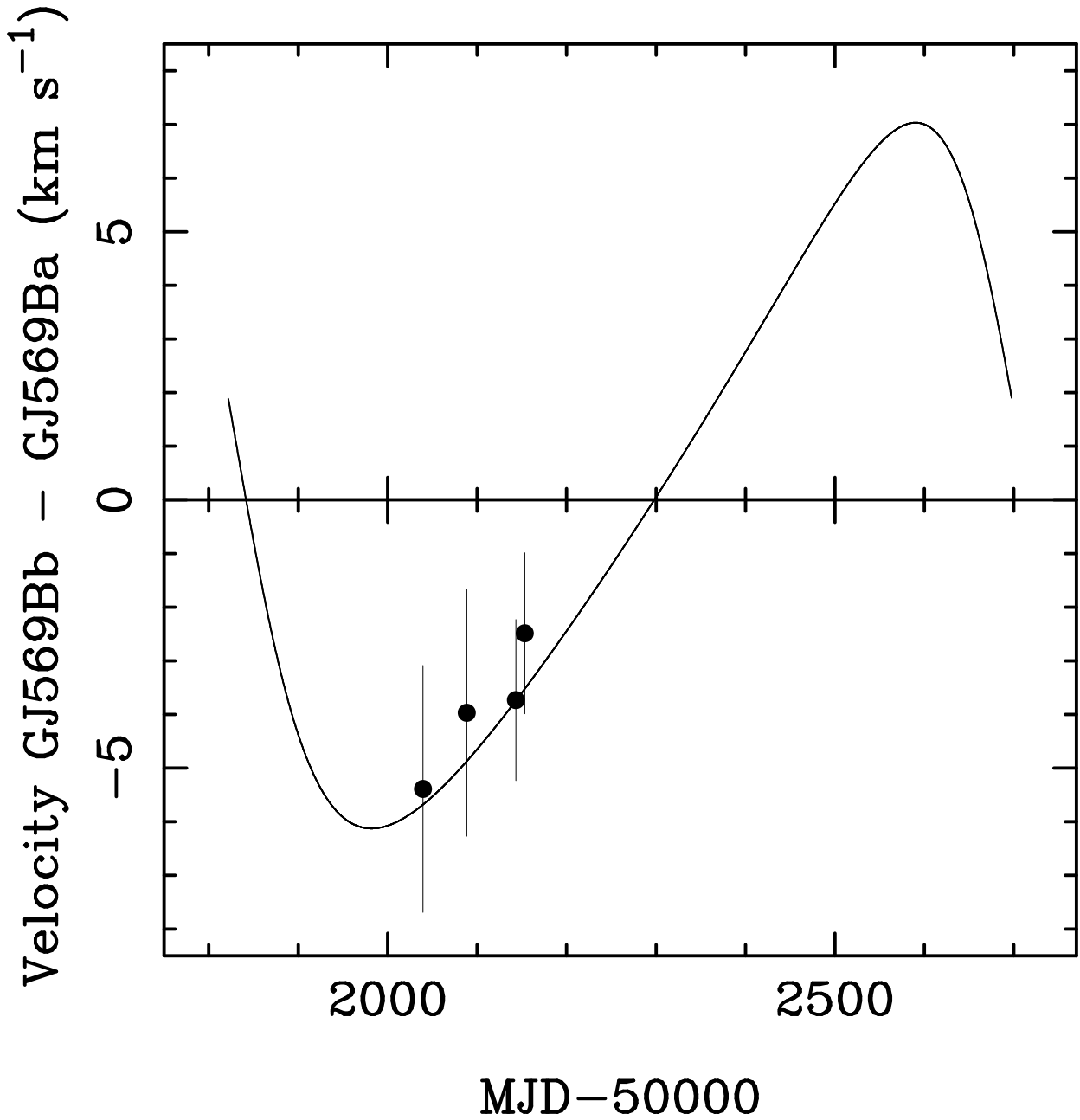}
\caption{Relative velocity of GJ\,569\,Bb with respect GJ\,569\,Ba plotted as a function of the epoch of the observations. Overplotted onto the data is the predicted velocity curve for the known orbital parameters and total mass of the pair. \label{veloc1}}
\end{figure}

The relative velocity of GJ\,569\,Bb with respect to Ba (or vice versa) is independent of any other velocity. It is only a function of the orbital parameters and the total mass of the pair. We have plotted in Fig.~\ref{veloc1} our velocities (solid circles) against the epoch of the observations. Overplotted is the velocity curve computed for the updated orbit of Table~\ref{orbit}. The agreement between the spectroscopic and astrometric data is excellent (within 1$\sigma$ the observational error bars), confirming the good quality of the Keplerian solution. 

The mass of each member of the binary is related to the radial velocity by the following equations:
% \begin{equation}
% K\,=\,\frac{2\pi\,a\,{\rm sin}\,i}{P\,(1-e^2)^{1/2}\,(M_{Ba}+M_{Bb})}
% \end{equation}
% \begin{equation}
% v_{Ba}\,=\,v_\circ\,+\,K\,M_{Bb}\,[{\rm cos}\,(\nu+\omega)\,+\,e\,{\rm cos}\,\omega]
% \end{equation}
% \begin{equation}
% v_{Bb}\,=\,v_\circ\,-\,K\,M_{Ba}\,[{\rm cos}\,(\nu+\omega)\,+\,e\,{\rm cos}\,\omega]
% \end{equation}
\begin{equation}
K\,=\,\frac{2\pi\,a\,{\rm sin}\,i}{P\,(1-e^2)^{1/2}\,[M_{Ba}\,(1\,+\,q)]}  \label{eq1}
\end{equation}
\begin{equation}
v_{Ba}\,=\,v_\circ\,+\,K\,q\,M_{Ba}\,[{\rm cos}\,(\nu+\omega)\,+\,e\,{\rm cos}\,\omega]  \label{eq2}
\end{equation}
\begin{equation}
v_{Bb}\,=\,v_\circ\,-\,K\,M_{Ba}\,[{\rm cos}\,(\nu+\omega)\,+\,e\,{\rm cos}\,\omega]  \label{eq3}
\end{equation}
where $P$, $i$, $a$, $w$, $e$ are the orbital parameters as given in Table~\ref{orbit}, $\nu$ is the true anomaly, and the velocity and mass of GJ\,569\,Ba are indicated by $v_{Ba}$ and $M_{Ba}$, respectively. Equations \ref{eq1} to \ref{eq3} are formulated using the mass ratio, $q$\,=\,$M_{Bb}$/$M_{Ba}$, of the binary. The quantity $v_\circ$ accounts for the long-term value of the radial velocity of the centre of mass of the binary, which includes the radial velocity of the triple system, the orbital motion of the centre of mass of the pair around GJ\,569\,A, and the average velocity of the binary about its barycentre. The total mass of GJ\,569\,Bab and the shape of the velocity curves of each component are fixed by the astrometric orbit; hence, there are only two variables in equations (2) and (3): $M_{Ba}$ and $v_\circ$. 

\begin{figure}
\epsscale{1.0}
\plotone{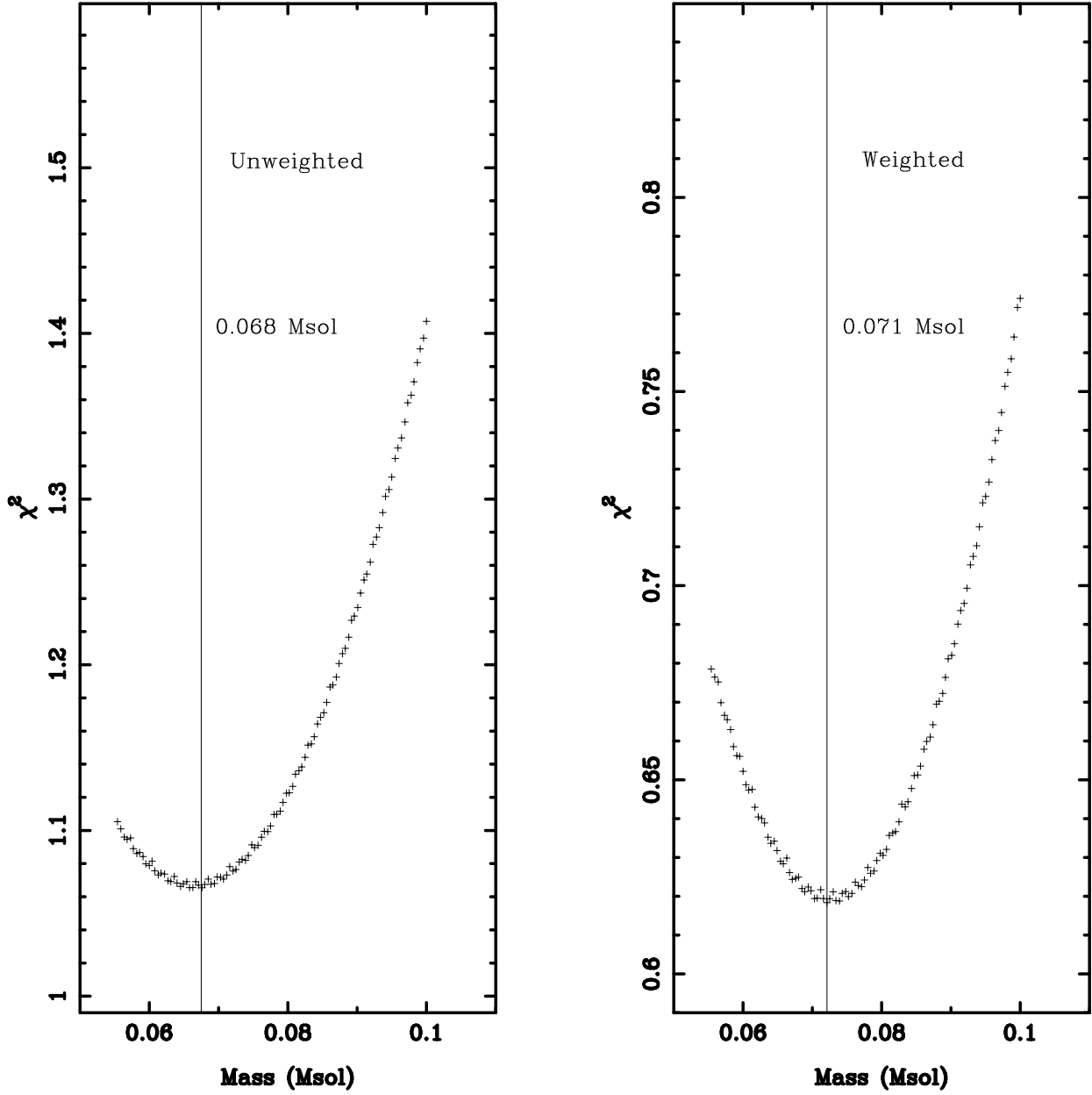}
\caption{The $\chi^2$ as a function of possible masses of GJ\,569\,Ba. For each mass, the systemic velocity of the pair has also been minimized. The unweighted ({\sl left panel}) and weighted ({\sl right panel}) heliocentric velocities of GJ\,569\,Ba and Bb are best reproduced by the mass corresponding to the minimum of the $\chi^2$ (vertical solid lines). The mass step in the calculations is 5$\times$10$^{-4}$\Msol. \label{chi2}}
\end{figure}

We have produced a grid of pairs of computed velocity curves for $M_{Ba}$ and $v_\circ$ values in the range 0.055--0.100\Msol~and from $-$9 to $-$15\,km\,s$^{-1}$, respectively, and have compared these velocities to our observed heliocentric data of Table~\ref{vrad}. We note that 0.055\Msol~is the minimum possible mass of GJ\,569\,Ba (at 3$\sigma$) according to purely astrometric and photometric considerations (Section~\ref{improvedorbit}). The upper limit to the mass has been arbitrarily imposed. The trial long-term velocity interval covers the range between the largest and smallest heliocentric velocity of the components. To find the optimal mass and long-term velocity we have minimized the chi-squared ($\chi^2$) function given by the following equation:
\begin{equation}
\chi^2\,=\,\Sigma\,\frac{[v\,-\,v_i]^2}{\sigma_i^2}
\end{equation}
where $\sigma_i$ stands for the errors in the velocity measurements. This is, observed velocities ($v_i$) were weighted by their corresponding uncertainties so that the best quality measurements are more important in finding the best fit. Unweighted heliocentric velocities were also compared to the computations by imposing $\sigma_i$\,=\,1\,km\,s$^{-1}$. Figure~\ref{chi2} shows the resulting $\chi^2$ against mass, and Table~\ref{bestfit} summarizes the best fits. The best-fit radial velocity curves correspond to the most likely masses of 0.071\Msol~(GJ\,569\,Ba), 0.054\Msol~(GJ\,569\,Bb) for the weighted data, and 0.068\Msol~(GJ\,569\,Ba), 0.057\Msol~(GJ\,569\,Bb) for the unweighted data. We note, however, that six radial velocity measurements (three per object) are probably too few to accurately constrain the systemic velocity of the binary. Consequently, the uncertainty of the parameters including the covariance of $v_\circ$ and $M_{Ba}$ is rather large, being $\pm$1.8\,km\,s$^{-1}$ and $\pm$0.052\Msol, respectively, obtained by increasing the $\chi^2$ by 2.3. Dispersions of $\pm$0.45\,km\,s$^{-1}$ and $\pm$0.011\Msol~(given in Table~\ref{bestfit}) are derived from the various $\chi^2$ solutions found when changing the astrometric orbital parameters by their corresponding uncertainties (given in Table~\ref{orbit}). This mass uncertainty is also consistent with the discussion below. 

There is another way to constrain the mass ratio of the binary. It makes use of the luminosity and $T_{\rm eff}$ of each component. We will update the luminosities of Paper~I by using the most recent $J$-band bolometric corrections given by Dahn et al$.$ \cite{dahn02}. GJ\,569\,Ba and Bb were assigned updated luminosities of log\,$L/L_\odot$\,=\,$-$3.35\,$\pm$\,0.07 and $-$3.58\,$\pm$\,0.07, respectively. By applying the astrophysical definition of effective temperature, we have obtained the radii of the components and found them to be: $R_{Ba}$\,=\,0.119\,$\pm$\,0.020\,$R_{\odot}$ and $R_{Bb}$\,=\,0.102\,$\pm$\,0.020\,$R_{\odot}$. The uncertainty in the radii has been derived from the errors of $\pm$0.07\,dex and $\pm$100\,K in luminosity and $T_{\rm eff}$, respectively. In Sect$.$~\ref{synthetic} we have proved that the surface gravities of GJ\,569\,Ba and Bb are quite alike, implying that the mass ratio of the pair, $q$, is directly related to the square ratio of the radii. Thus, it follows that $q$\,=\,0.74\,$\pm$\,0.25. This result is indeed very similar to that obtained from the radial velocity fitting (Table~\ref{bestfit}), suggesting that the long-term velocity of GJ\,569\,Bab and the masses of the components are reliable to some extent.

As discussed in Sect$.$~\ref{improvedorbit}, the 3\,$\sigma$ upper limit of the mass of GJ\,569\,Bb is 0.070\Msol. At the 99\%~confidence level, the dynamical mass of GJ\,569\,Bb is in the interval 0.034--0.070\Msol, and the dynamical mass of GJ\,569\,Ba is in the range 0.055--0.087\Msol. These intervals nicely overlap with the mass estimates listed in Table~5 of Paper~I, which were based solely on the comparison of the photometry with evolutionary models. More radial velocity measurements are needed to decrease the mass uncertainty and detect the presence of much tiny objects orbiting any of the brown dwarfs. This sets the mass ratio of the binary, $q$, to be $>$0.4, being the most likely value $q$\,=\,0.75--0.85. The most conservative location of the hydrogen burning mass limit, i.e., the borderline between stars and brown dwarfs, is 0.072\Msol~for solar metallicity (Chabrier \& Baraffe~\cite{chabrier00}). The dynamical mass of GJ\,569\,Bb is definetively below the substellar frontier, confirming it as the {\sl first genuine brown dwarf known independently of models}. GJ\,569\,Ba is very likely a brown dwarf too, although it may lie at the star--brown dwarf borderline.

\section{Discussion}

For the first time, it is possible to evaluate the predictions of theoretical evolutionary models in the substellar regime, by comparing isochrones to the dynamical mass, $T_{\rm eff}$, gravity and luminosity of GJ\,569\,Ba and Bb. For this independent test, we need to constrain the age of the triple system. On the other hand, GJ\,569\,ABab may provide a calibration of the brown dwarf--mass--age--luminosity relation.

\subsection{The age of GJ\,569\,ABab}

The membership of GJ\,569\,ABab to a stellar kinematic group would help us set limits upon the age of the system. The space velocity components, $UVW$, of GJ\,569\,A put it firmly in the young disk region of the space motion groups as defined by Eggen \cite{eggen69}. Furthermore, these components can be compared to the space motion of the Ursa Major (UMa) moving group (core and extended halo, sometimes called the Sirius supercluster) and the new supercluster found by Chereul, Cre\'ez\'e, \& Bienaym\'e \cite{chereul99}, which has a mean velocity between the Sun and the Sirius supercluster. 

In a recent paper, King et al$.$ \cite{king03} used the radial velocity of GJ\,569\,A ($-7.17$\,$\pm$\,0.28\,km\,s$^{-1}$) measured by Marcy et al$.$ \cite{marcy87} and the Hipparcos parallax and proper motion of GJ\,569\,A to study its Galactic velocity components. These authors concluded that the membership of GJ\,569\,A in the young UMa moving group is uncertain. Using the same data, Kenworthy et al$.$ \cite{kenworthy01} argued that the kinetics of GJ\,569\,A is within 7.2\,km\,s$^{-1}$ of the space motion of UMa. Nevertheless, none of these authors took into account the orbital motion of GJ\,569\,A about the barycenter of the triple system. The mass of this M2.5V-class star is estimated at 0.3--0.4\Msol~(Gorlova et al$.$ \cite{gorlova03}). By applying simple Keplerian equations, we inferred that the observed radial velocity of GJ\,569\,A may be affected by an uncertainty of up to $\pm$\,1.3\,km\,s$^{-1}$. Hence, the velocity vector of GJ\,569\,A and its 1\,$\sigma$ error bar turn out to be $U$\,=\,7.8\,$\pm$\,0.7, $V$\,=\,3.3\,$\pm$\,0.3, $W$\,=\,$-$13.3\,$\pm$\,1.1\,km\,s$^{-1}$. We have used the mathematical formulas of Johnson \& Soderblom \cite{johnson87} to calculate the space-velocity components and their uncertainties from the errors in the observational quantities (parallax, proper motion and radial velocity). Our result is in very much agreement with that of Kenworthy et al$.$ \cite{kenworthy01} and King et al$.$ \cite{king03}, except for our larger error bars that mainly come from the uncertainty in the orbital velocity of GJ\,569\,A. 

From the literature, we have obtained that the mean Galactic space-velocity components of the UMa moving group are the following: $U$\,=\,12.8, $V$\,=\,2.2, $W$\,=\,$-$8.6\,km\,s$^{-1}$ (Eggen \cite{eggen92}; Soderblom \& Mayor \cite{soderblom93}; Orlov et al$.$ \cite{orlov95}; Chen et al$.$ \cite{chen97}; Asiain et al$.$ \cite{asiain99}; Chereul et al$.$ \cite{chereul99}; Montes et al$.$ \cite{montes01}; King et al$.$ \cite{king03}). The velocity dispersion of $UVW$ among individual stars of the UMa group is typically around 3\,km\,s$^{-1}$ (Madsen, Dravins, \& Lindegren \cite{madsen04}), although dispersions as high as 7\,km\,s$^{-1}$ are seen in various works. This large scatter is not surprising since the group is unbound and spatially extended in the Galactic disk. The space motion vector of GJ\,569\,A overlaps with that of the UMa group if the uncertainties are multiplied by a factor of 2, which indicates that its membership in the UMa cluster is possible, but not very probable. In the kinematic planes shown in Fig$.$ 17 of Chereul et al$.$ \cite{chereul99}, GJ\,569\,A is positioned to the left of the Sirius supercluster, very close to the location of the new supercluster reported by these authors. The $UVW$ coordinates of GJ\,569\,A differ at only the 1\,$\sigma$ level with respect the space motion of this new supercluster, in particular with the stream moving at $U$\,=\,5.3\,$\pm$\,2.7, $V$\,=\,6.5\,$\pm$\,3.6, $W$\,=\,$-$11.4\,$\pm$\,4.0\,km\,s$^{-1}$ (stream 2-43 according to the nomenclature of the discoverers). On the basis of its space velocity, GJ\,569 may kinematically belong to any of these two moving groups.

The age of the UMa cluster has been widely discussed in the literature. The canonical quoted age of the nucleus is around 300\,Myr (Soderblom et al$.$ \cite{soderblom_stauffer93}, and references therein), although some older values (up to 600\,Myr) are claimed more recently (King et al$.$ \cite{king03}, and references therein). The massive stars of the new supercluster of Chereul et al$.$ \cite{chereul99} have ages very similar to the Sirius supercluster (100--1000\,Myr). Nevertheless, based on Str\"omgren photometry, the stream 2-43 contains rather young stars, with ages in the interval 300--800\,Myr. Hence, we conclude that if GJ\,569 is a member of any of these kinematic groups, its age must be constrained in the range 300--800\,Myr. 

The spectroscopic and photometric properties of the star GJ\,569\,A also support the youth of the system. The age of GJ\,569\,A has been deeply studied in the literature (e.g., Forrest et al$.$ \cite{forrest88}; Henry \& Kirkpatrick \cite{henry90}; Mart\'\i n et al$.$ \cite{martin00}; Kenworthy et al$.$ \cite{kenworthy01}; Gizis, Reid, \& Hawley \cite{gizis02}). The various groups conclude that GJ\,569\,A exhibits many of the attributes of young stars (chromospheric activity, flaring), and claim that the likely age of GJ\,569\,A is in the interval 120--1000\,Myr. In addition, the color-magnitude diagrams shown in Paper~I, which included the decomposition of the near-infrared photometry of the binary, also suggest that GJ\,569 is a young system.

\subsection{Mass--luminosity and mass--$T_{\rm eff}$ relations \label{comparisons}}

We note that our $T_{\rm eff}$ values from Paper~I are within 50\,K of those recently found by Gorlova et al$.$ \cite{gorlova03}, and within 75\,K of the new $T_{\rm eff}$--spectral type calibration by Dahn et al$.$ \cite{dahn02}. The updated luminosities of Sect$.$~\ref{mass_determination} are also within the error bars of previous estimates.

\begin{figure*}
\epsscale{1.0}
\plottwo{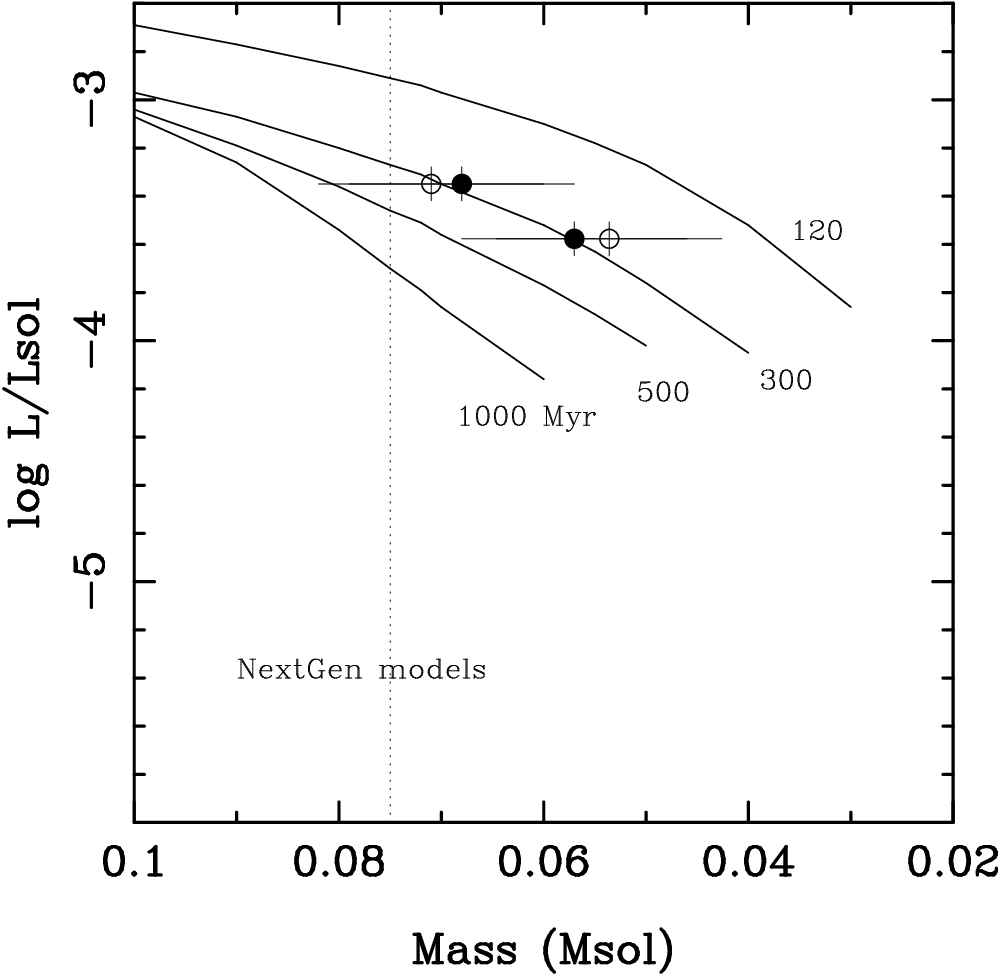}{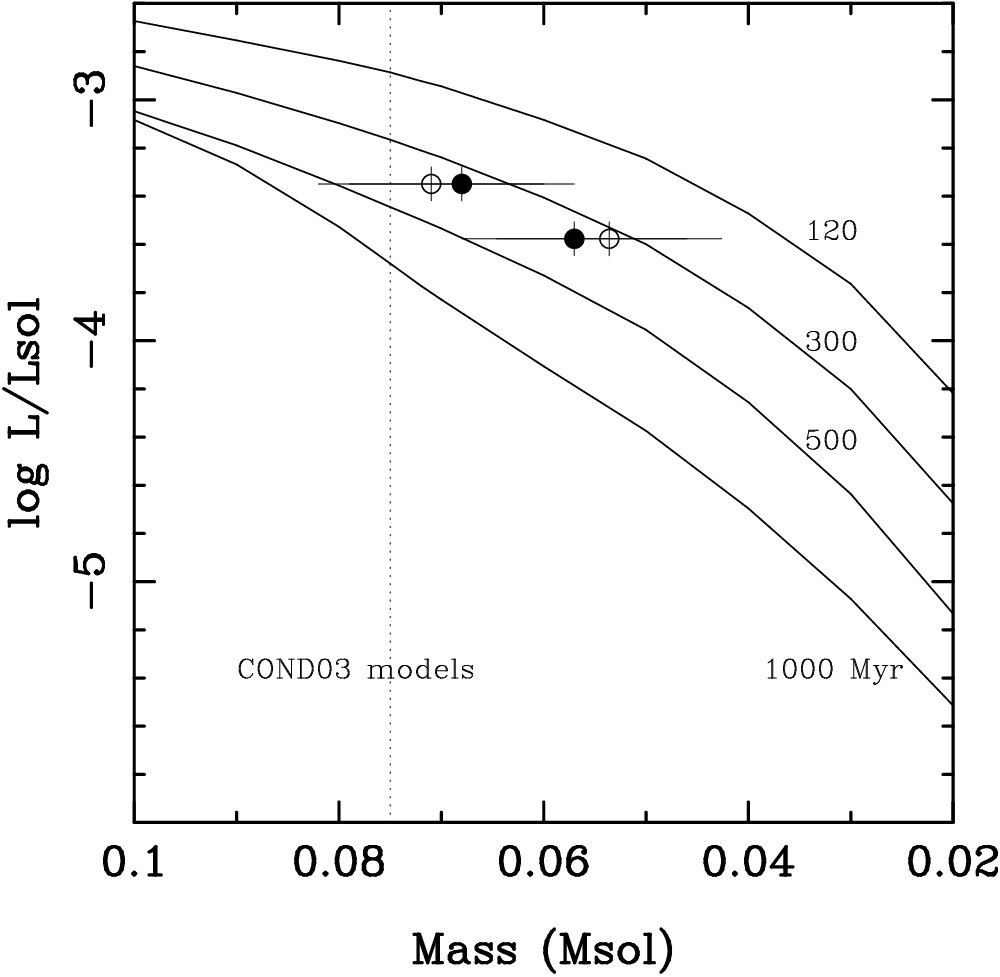}
\plottwo{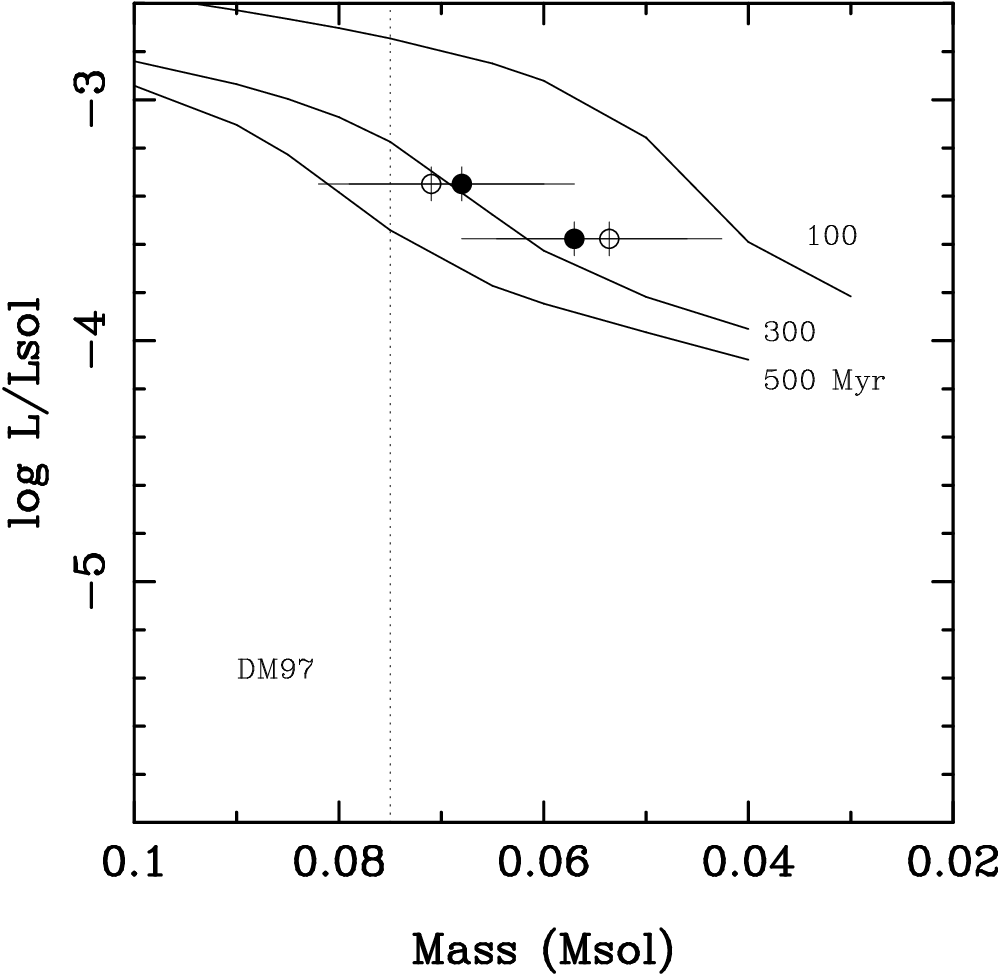}{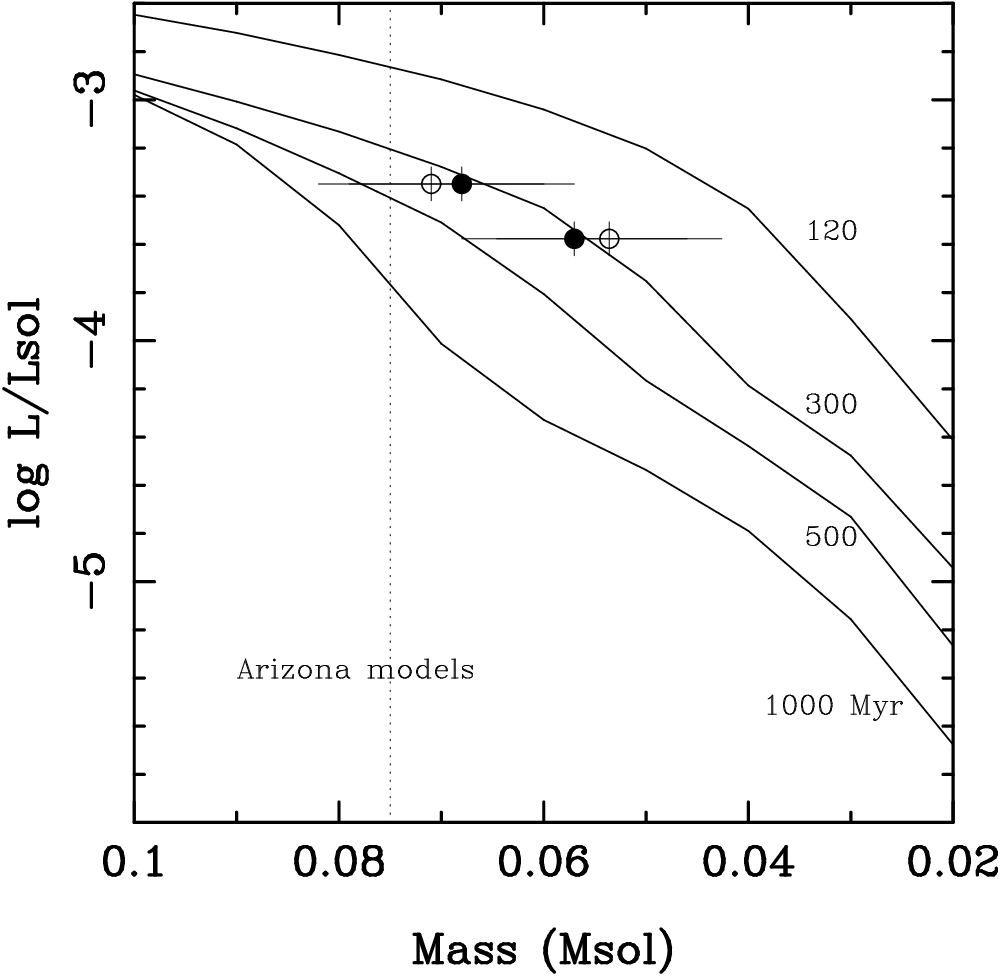}
\caption{Mass-luminosity relationships from various groups (NextGen models of Baraffe et al$.$ \cite{baraffe98} --- top left; COND models of Baraffe et al$.$ \cite{baraffe03} --- top right; D'Antona \& Mazzitelli \cite{dantona97} models --- bottom left; Arizona models of Burrows et al$.$ \cite{burrows97} --- bottom right). The weighted (open circles) and unweighted (solid circles) masses of GJ\,569\,Ba and Bb are also plotted together with 1\,$\sigma$ error bars. The vertical dotted line represents the substellar limit. Isochrones are labelled with ages in Myr. \label{mass_lum}}
\end{figure*}

Figures~\ref{mass_lum} and~\ref{mass_teff} show GJ\,569\,Ba and Bb plotted on the mass--luminosity and mass--$T_{\rm eff}$ diagrams, respectively. Both the weighted and unweighted dynamical masses are displayed. For the figures, we have considered four different sets of pre-main sequence evolutionary models: the NextGen models of Baraffe et al$.$ \cite{baraffe98}, the corrected isochrones of D'Antona \& Mazzitelli \cite{dantona97}, hereafter DM97, the ``Arizona'' tracks of Burrows et al$.$ \cite{burrows97}, and the most recent release of the {\sc cond} models by Baraffe et al$.$ \cite{baraffe03}. These include essentially all of the models widely used by many authors for substellar studies. The two evolutionary tracks of Baraffe et al$.$ (1998, 2003) are usually referred to as the ``Lyon'' models. For a detailed description of the evolutionary models, see the corresponding authors. Nevertheless, we note that all of them are state-of-the-art calculations that employ sophisticated physiscs and boundary conditions. Since the two brown dwarfs of GJ\,569\,Bab are supposed to be coeval, a single isochrone is expected to fit both components simultaneously. Models must also reproduce the absolute and relative location of the brown dwarfs in the diagrams.

From Figs$.$~\ref{mass_lum} and~\ref{mass_teff}, the four sets of isochrones indicate that GJ\,569\,Bab is relatively young, $\sim$200--1000\,Myr, with a likely age at around 300\,Myr. We note that this result agrees with the discussion on the system age of the previous section. It is important to remark that all models yield reasonable fits to the observed parameters of the brown dwarf binary only if the system turns out to be about a few hundred million years old. In the following, we will briefly discuss on the fine details of the comparisons.

We recall that the masses of GJ\,569\,Ba and Bb have been derived independently of models, whereas temperatures and luminosities hinge to some extent on external calibrations. For the Lyon and Arizona models, both the primary and secondary components of the binary lie within 1\,$\sigma$ of the 300-Myr isochrones depicted in Fig$.$~\ref{mass_lum}. We are now considering the unweighted dynamical masses. However, the luminosity prediction of the 300-Myr DM97 track for the dynamical mass of GJ\,569\,Bb deviates by 2\,$\sigma$. In general, the Lyon and Arizona theoretical mass--luminosity relations for ages around a few hundred million years agree acceptably well with our observations of the binary GJ\,569\,Bab at the level of 1\,$\sigma$ the observational uncertainties.

Larger differences are found when comparing GJ\,569\,Ba and Bb to the mass--$T_{\rm eff}$ relations shown in Fig$.$~\ref{mass_teff}. While the Lyon models at the age of 300\,Myr appear to reproduce within 1\,$\sigma$ the slope formed by the members of the pair, GJ\,569\,Ba and Bb lie at different isochrones of the Arizona models. The secondary and primary appear between 1 and 2\,$\sigma$ up and down, respectively, of the predictions of the 300\,Myr DM97 isochrone. 

\subsection{Model atmospheres and evolutionary isochrones}

\begin{figure*}
\epsscale{1.0}
\plottwo{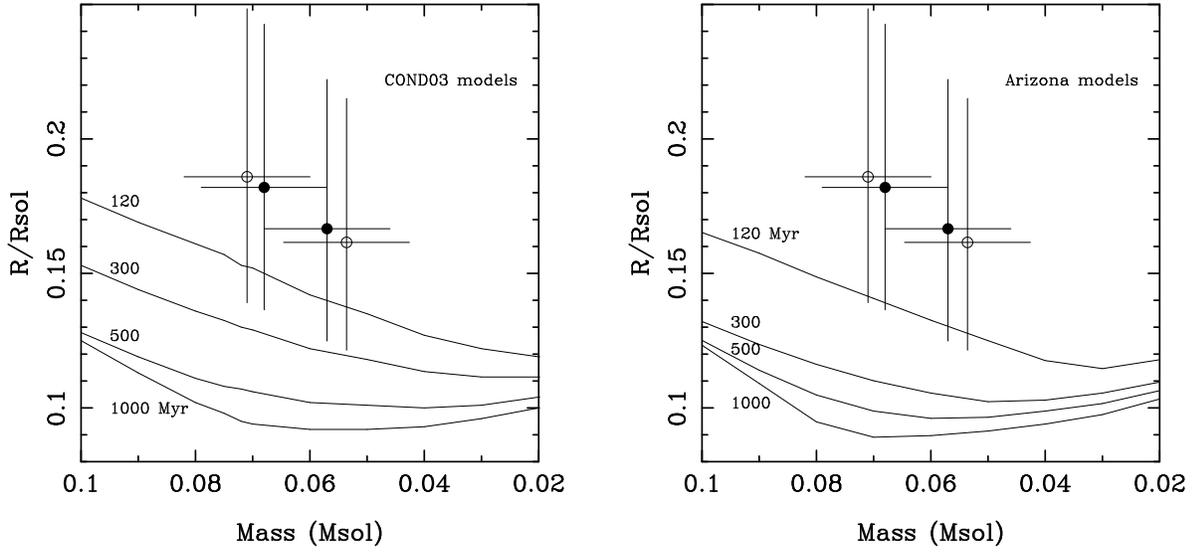}{f12b.ps}
\caption{Radius vs$.$ mass for the models of Baraffe et al$.$ \cite{baraffe03} {\sl (left panel)} and Burrows et al$.$ \cite{burrows97} {\sl (right panel)}. The weighted (open circles) and unweighted (solid circles) masses of GJ\,569\,Ba and Bb are also plotted together with 1\,$\sigma$ error bars. The radii of these objects have been obtained from log\,$g$\,=\,4.75\,$\pm$\,0.25, which results from the synthetic spectral fitting.  \label{radii}}
\end{figure*}

For a test of consistency between evolutionary models, model atmospheres and $T_{\rm eff}$ calibrations, we have obtained the surface gravity of GJ\,569\,Ba and Bb by means of synthetic spectral fitting (see Sect$.$~\ref{synthetic}). Radii can be derived from log\,$g$ and the Newton's law for Gravitation. Adopting $g_\odot$\,=\,27380.215\,cm\,s$^{-2}$ and the unweighted dynamical masses, the radius of GJ\,569\,Ba and Bb is $R/R_\odot$\,=\,0.18\,$^{+0.06}_{-0.05}$ and 0.17\,$^{+0.06}_{-0.04}$, respectively. These values are plotted against mass in Fig$.$~\ref{radii}, together with the Lyon and Arizona isochrones. There is a large discrepancy (about a factor of 1.6) between the predictions of the evolutionary models and the radii obtained from the spectral fitting. Such a discrepancy translates into a difference in log\,$g$ of the order of 0.5\,dex. Very recently, Mohanty et al$.$ \cite{mohanty04} have reported finding very low gravities in young late-type stars of the Upper Scorpius association as compared to theoretical evolutionary model predictions. According to these authors, the differences are the largest for the coolest types ($\ge$M7).

We note, however, that the radii derived in Sect$.$~\ref{mass_determination} are in much better agreement with the evolutionary models if the age of the pair is about a few hundred million years.

The discrepancy between the radii predictions for GJ\,569\,Ba and Bb and those inferred from the spectroscopically derived surface gravities could be due to a combination of the uncertainties in the $T_{\rm eff}$--spectral type calibrations, the spectral synthesis, the model atmospheres, the log\,$g$ step employeed in the synthetic calculations, and the evolution isochrones (see Mohanty et al$.$ \cite{mohanty04} for a rather extensive discussion). On the other hand, the formation processes and evolution of binaries are not well understood, and tracks describing the evolution of ``single'' objects might not be valid. Nevertheless, warmer values of the $T_{\rm eff}$ assignement of GJ\,569\,Ba and Bb would bring to a better harmony the radii predictions of evolutionary models and spectral fitting techniques. From our spectral fitting analysis, we observed that larger best-fit values of log\,$g$ are found for hotter temperatures (see the contours of Fig$.$ \ref{minima}), e.g., 2600\,K and log\,$g$\,=\,5.0 (collisions with H and H$_2$) and 5.5 (collisions with H), although they do not minimize the function defined by Pavlenko et al$.$ \cite{pav04}. A temperature of 2600\,K is marginally consistent with the most recent calibrations for M8.5--M9 dwarf objects. On the other hand, an increase of 150\,K in the $T_{\rm eff}$ of GJ\,569\,Ba and Bb does not change significantly the discussion of Sect$.$~\ref{comparisons}. Actually, the Arizona models would fit better the observed parameters of the binary at 300\,Myr, and the Lyon models would predict a slightly younger age for the system. 

In any case, the spectral fitting of the K\,{\sc i} doublet at 1.25\,$\micron$ and the radii derivation of Sect$.$~\ref{mass_determination} suggest that the radii of GJ\,569\,Ba and Bb are larger than the evolutionary model predictions for the oldest ages ($\sim$1\,Gyr). This might be due to the young age of the objects (the brown dwarfs would be contracting); on the other hand, part of this discrepancy might be real. Similar discrepancies have been observed in low-mass eclipsing binaries and in the interferometric radii measurements of M-dwarfs as reported by various groups. For example, models fail to predict the radius of the transiting planet HD\,209458\,b (predictions are too small, Baraffe et al$.$ \cite{baraffe03}), and underestimate the size of cool stars up to 10\%~(e.g., Lane, Boden, \& Kulkarni \cite{lane01b}; Ribas \cite{ribas03}; S\'egransan et al$.$ \cite{segransan03}). For a definitive and critical constraint of substellar model predictions, the size of each brown dwarf of the binary GJ\,569\,Bab should be determined by means that do not include any theoretical calculations or external calibrations, e.g., near infrared interferometry, or photometric monitoring to determine rotation periods and infer radii by assuming that the rotation angles are perpendicular to the orbital plane of the binary.

\subsection{Final remarks}

The mass ratio of GJ\,569\,Bab is $q$\,$>$\,0.4, very likely $q$\,=\,0.75--0.85 (Table~\ref{bestfit}). Many other binary brown dwarfs have similar mass ratio estimates (based on the comparison to evolutionary models), like, e.g., PPl\,15 ($q$\,=\,0.87, Basri \& Mart\'\i n \cite{basri99}), 2MASS\,J0231016--040618 ($q$\,=\,0.7, Close et al$.$ \cite{close02a}), 2MASS\,J1426316+155701 ($q$\,=\,0.9, Close et al$.$ \cite{close02b}), and HD\,130948BC ($q$\,=\,0.9, Potter et al$.$ \cite{potter02}). Close et al$.$ \cite{close03} presents the compilation of nearly all very low mass, field resolved binary systems known to date, concluding that they tend to have nearly equal mass components ($q$\,$\sim$\,0.9). Furthermore, Mart\'\i n et al$.$ \cite{martin03} have investigated the binary fraction among young brown dwarfs of the Pleiades cluster (120\,Myr), finding that the frequency of these binaries is about 15\%~(for separations between 7 and 12\,AU), with a clear trend to high mass ratios ($q$\,$>$\,0.7). Nevertheless, the discovery of substellar pairs like $\epsilon$\,Indi\,B ($q$\,=\,0.6, McCaughrean et al$.$ \cite{mccaughrean04}) suggests that binaries with moderate mass ratios are also possible, albeit they appear to be less frequent. Any model describing the formation mechanisms of very low mass stars and brown dwarfs should be capable of explaining these observations.

\section{Summary and conclusions}

Using the Keck\,II Adaptive Optics facility and the NIRSPEC instrument, we have obtained new near-infrared images that have allowed us to improve the astrometric solution of the substellar binary GJ\,569\,Bab, which is a proper motion companion to the M2.5V-type star GJ\,569\,A. The improved orbital parameters lie within 1-$\sigma$ the previously published values of Lane et al$.$ \cite{lane01a}, suggesting that the orbit is now determined to a high degree of accuracy. The total mass of the pair turns out to be 0.125\,$\pm$\,0.015\Msol~at the 3-$\sigma$ confidence level.

We have also obtained high-resolution spectra of each component, GJ\,569\,Ba and Bb, using the echelle mode of NIRSPEC on four different occasions during one orbital period. The spectra, with a resolution of 0.55\,\AA~($R$\,$\sim$\,22500), cover the wavelength range between 1.2400 and 1.2575\,$\mu$m, which includes the K\,{\sc i} doublet at 1.2436 and 1.2526\,$\mu$m. We have used these spectroscopic data to measure radial velocities of each member of the binary system with an accuracy between 1.1 and 2\,km\,s$^{-1}$. The combination of the astrometric orbit and the spectroscopic velocities has allowed us to derive the dynamical masses of GJ\,569\,Ba and Bb and the systemic velocity of the pair by means of a $\chi^2$ fitting technique. We note that our results show a marked correlation, which yields relatively large uncertainties in the parameters. More radial velocities measurements are needed to break the covariance problem. Holding the long-term velocity of GJ\,569\,Bab fixed at its best-fit values, the masses obtained are $M_{GJ\,569\,Ba}$\,=\,0.068 and $M_{GJ\,569\,Bb}$\,=\,0.057\Msol~for the unweighted velocities, and $M_{GJ\,569\,Ba}$\,=\,0.071 and $M_{GJ\,569\,Bb}$\,=\,0.054\Msol~for the weighted data. This implies that the mass ratio of the binary is greater than 0.4, being the most likely value $q$\,=\,0.75--0.85. The dynamical masses are affected by an uncertainty of 16--20\%~(the error increases if the covariance problem is taken into account). Because of this uncertainty, the more massive component, GJ\,569\,Ba may lie at the star--brown dwarf borderline. However, the mass of the secondary, GJ\,569\,Bb, is very likely below the substellar limit. For the first time, the substellar mass of one brown dwarf is dynamically confirmed, i.e., without any theoretical assumption.

The photometric and spectroscopic properties of the primary star GJ\,569\,A, as well as its $UVW$ space velocities, suggest that the age of the triple system is in the interval 300--800\,Myr. The dynamical masses of GJ\,569\,Ba and Bb, along with their effective temperatures and luminosities, have been compared to the mass-luminosity and mass-$T_{\rm eff}$ relations given by four different sets of state-of-the-art evolutionary models. We have found that these models reasonably reproduce the absolute and relative locations of GJ\,569\,Ba and Bb in the luminosity vs$.$ mass and $T_{\rm eff}$ vs$.$ mass diagrams if the age of the binary is about a few hundred million years. Hence, as for luminosities and $T_{\rm eff}$'s, evolutionary models appear to exhibit good performance in the regime of high substellar masses.

The observed high-resolution spectra of GJ\,569\,Ba (M8.5V) and Bb (M9V) have been compared to various spectral synthesis using three different sets of model atmospheres. Our spectral fitting procedure yields $T_{\rm eff}$'s that are very similar (within 100\,K) to the calibrations for M8--M9-class objects available in the literature, and surface gravities that are quite small as compared to the predictions of evolutionary models. A similar discrepancy has been recently reported by Mohanty et al$.$ \cite{mohanty04}. We argue that slightly warmer values of the $T_{\rm eff}$'s of the binary components would bring to a better agreement the gravity predictions of evolutionary models and spectral synthesis.

\acknowledgements

We thank Barbara Schaefer, Randy Cambell and David Le Mignant for their help with the 2001 Sep 01 Keck service observations. We also thank F$.$ Allard and T$.$ Tsuji for providing machine-readable versions of their model atmospheres. We thank the anonymous referee for his/her comments. Data presented herein were obtained at the W.\,M$.$ Keck Observatory, which is operated as a scientific partnership between the California Institute of Technology, the University of California, and NASA. The Observatory was made possible by the generous financial support of the W.\,M$.$ Keck Foundation. The authors wish to extend special thanks to those of Hawaiian ancestry on whose sacred mountain we are privileged to be guests. This research has made use of the Simbad database, operated at CDS, Strasbourg, France. YP acknowledges partial financial support from PPARC and the Royal Society and Small Research Grant from American Astronomical Society. BFL acknowledges support from a Pappalardo Fellowship in Physics. This work was partly carried out with support from the MCyT projects AYA2003-05355 and Ram\'on y Cajal. 

Facilities: \facility{Keck(NIRSPEC)}.

\clearpage

\clearpage

\clearpage

\clearpage

\clearpage

\clearpage

\clearpage

\clearpage

\clearpage

\clearpage

\begin{figure}
\epsscale{1.0}
\plotone{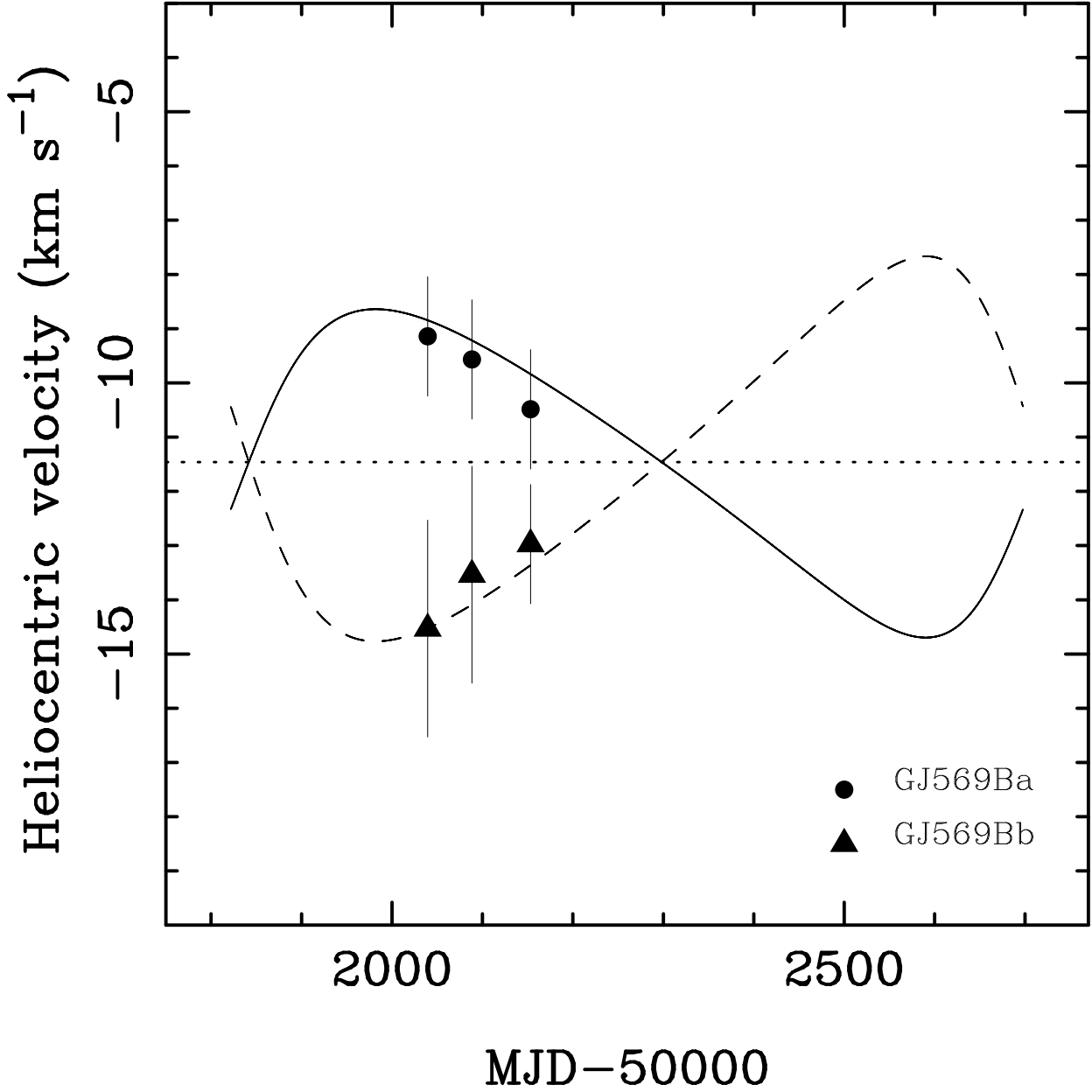}
\caption{Heliocentric velocities of GJ\,569\,Ba (circles) and Bb (triangles), together with the best-fit velocity curves (solid and dashed lines). The dotted, horizontal line indicates the centre of mass velocity of the pair. \label{veloc2}}
\end{figure}

\clearpage

\clearpage

\begin{figure*}
\epsscale{1.0}
\plottwo{f11a.ps}{f11b.ps}
\plottwo{f11c.ps}{f11d.ps}
\caption{Mass-$T_{\rm eff}$ relationships from various groups (NextGen models of Baraffe et al$.$ \cite{baraffe98} --- top left; COND03 models of Baraffe et al$.$ \cite{baraffe03} --- top right; D'Antona \& Mazzitelli \cite{dantona97} models --- bottom left; Arizona models of Burrows et al$.$ \cite{burrows97} --- bottom right). The weighted (open circles) and unweighted (solid circles) masses of GJ\,569\,Ba and Bb are also plotted together with 1\,$\sigma$ error bars. The vertical dotted line represents the substellar limit. Isochrones are labelled with ages in Myr. \label{mass_teff}}
\end{figure*}

\clearpage

\clearpage

\begin{table}
\begin{center}
\caption{New astrometry of GJ\,569\,Bab.\label{astrometry}}
\bigskip
\begin{tabular}{lccr}
\tableline\tableline
\multicolumn{1}{l}{UT Date} &
\multicolumn{1}{c}{MJD} & 
\multicolumn{1}{c}{Separation} &
\multicolumn{1}{c}{P.A.}       \\
\multicolumn{1}{l}{    } &
\multicolumn{1}{c}{} & 
\multicolumn{1}{c}{(\arcsec) } &
\multicolumn{1}{c}{(deg)}      \\
\tableline
2001 Jun 28 & 52088.2913 & 0.1024\,$\pm$\,0.0012 & 352.6\,$\pm$\,2 \\
2001 Sep 01 & 52153.2211 & 0.1033\,$\pm$\,0.0010 &   9.7\,$\pm$\,2 \\
\tableline
\end{tabular}
\end{center}
\end{table}

\begin{table}
\begin{center}
\caption{Improved orbital parameters of GJ\,569\,Bab.\label{orbit}}
\bigskip
\begin{tabular}{lrclrcl}
\tableline\tableline
\multicolumn{1}{l}{Parameter} &
\multicolumn{3}{c}{This paper} & 
\multicolumn{3}{c}{Previous value\tablenotemark{a}} \\
\tableline
Total mass, $M$ (\Msol)   & 0.125 & $\pm$ & 0.005 & 0.123& $\pm$ & 0.008 \\
Period, $P$ (days)        & 876   & $\pm$ & 9     & 892  & $\pm$ & 25    \\
Eccentricity, $e$         & 0.32  & $\pm$ & 0.01  & 0.32 & $\pm$ & 0.02  \\
Semi-major axis, $a$ (AU) & 0.90  & $\pm$ & 0.01  & 0.90 & $\pm$ & 0.02  \\
Inclination, $i$ (deg)    & 34    & $\pm$ & 2     & 34   & $\pm$ & 3     \\
Arg$.$ periapsis, 				 
$\omega$ (deg)            & 257   & $\pm$ & 2     & 76\tablenotemark{b}   & $\pm$ & 4     \\
Long$.$ ascending				 
node, $\Omega$ (deg)      & 321.5 & $\pm$ & 2.0   & 141\tablenotemark{b}  & $\pm$ & 4     \\
Epoch (MJD), $T$ (days)   & 51822 & $\pm$ & 3     & 51820& $\pm$ & 4     \\
\tableline
\end{tabular}
\tablenotetext{a}{From Lane et al$.$ \cite{lane01a}.}
\tablenotetext{b}{There was an ambiguity of 180$^{\circ}$ in the determination of these two parameters in Lane et al$.$ \cite{lane01a}.}
\end{center}
\end{table}

\begin{table}
\begin{center}
\caption{Log of spectroscopic observations.\label{spectroscopy}}
\bigskip
\begin{tabular}{lcccc}
\tableline\tableline
\multicolumn{1}{l}{Object} &
\multicolumn{1}{c}{UT Date} &
\multicolumn{1}{c}{Exposure} &
\multicolumn{1}{c}{Airmass} &
\multicolumn{1}{c}{Weather} \\
\multicolumn{1}{l}{      } &
\multicolumn{1}{c}{    } &
\multicolumn{1}{c}{(s)     } &
\multicolumn{1}{c}{       } &
\multicolumn{1}{c}{       } \\
\tableline
GJ\,569\,Bab & 2001 May 10 & 3$\times$200  & 1.00--1.01 & Clouds \\  % HR 5567
GJ\,569\,Bab & 2001 Jun 28 & 2$\times$1200 & 1.02--1.04 & Clouds \\  % HR 5567
GJ\,569\,A   & 2001 Jun 28 & 2$\times$400  & 1.12--1.14 & Clouds \\  % HR 5567
GJ\,569\,Bab & 2001 Aug 22 & 3$\times$600  & 1.38--1.55 & Clouds \\  % HR 5931
GJ\,569\,Bab & 2001 Sep 01 & 2$\times$600  & 1.36--1.48 & Clear  \\  % SAO 101508
\tableline
\end{tabular}
\end{center}
\end{table}

\begin{table}
\begin{center}
\caption{Radial velocity measurements.\label{vrad}}
\bigskip
\begin{tabular}{cccc}
\tableline\tableline
\multicolumn{1}{c}{MJD} &
\multicolumn{1}{c}{$v$[GJ\,569\,Bb -- GJ\,569\,Ba]} & 
\multicolumn{1}{c}{$v_h$[GJ\,569\,Ba]} & 
\multicolumn{1}{c}{$v_h$[GJ\,569\,Bb]} \\
\multicolumn{1}{c}{} &
\multicolumn{1}{c}{(km\,s$^{-1}$)} & 
\multicolumn{1}{c}{(km\,s$^{-1}$)} & 
\multicolumn{1}{c}{(km\,s$^{-1}$)} \\
\tableline
52039.452 & $-$5.4\,$\pm$\,2.3 & $-$9.1\,$\pm$\,1.1  & $-$14.5\,$\pm$\,2.0 \\ % 10 May 2001
52088.314 & $-$4.0\,$\pm$\,2.3 & $-$9.6\,$\pm$\,1.1  & $-$13.5\,$\pm$\,2.0 \\ % 28 Jun 2001
52143.272 & $-$3.7\,$\pm$\,1.5 & \nodata             & \nodata             \\ % 22 Aug 2001
52153.240 & $-$2.5\,$\pm$\,1.5 & $-$10.5\,$\pm$\,1.1 & $-$13.0\,$\pm$\,1.1 \\ % 01 Sep 2001
\tableline
\end{tabular}
\end{center}
\end{table}

\clearpage

\begin{table}
\begin{center}
\caption{Best-fit masses of GJ\,569\,Ba and Bb.\label{bestfit}}
\bigskip
\begin{tabular}{ccccc}
\tableline\tableline
\multicolumn{1}{c}{ } &
\multicolumn{1}{c}{$v_\circ$} &
\multicolumn{1}{c}{$M$ (GJ\,569\,Ba)} &
\multicolumn{1}{c}{$M$ (GJ\,569\,Bb)} &
\multicolumn{1}{c}{$q$}               \\
\multicolumn{1}{c}{ } &
\multicolumn{1}{c}{(km\,s$^{-1}$)} &
\multicolumn{1}{c}{(\Msol)} &
\multicolumn{1}{c}{(\Msol)}  &
\multicolumn{1}{c}{ }                 \\
\tableline
Weighted   & $-$11.46\,$\pm$\,0.45 & 0.071\,$\pm$\,0.011 & 0.054\,$\pm$\,0.011 & 0.76\,$\pm$\,0.24 \\
Unweighted & $-$11.52\,$\pm$\,0.45 & 0.068\,$\pm$\,0.011 & 0.057\,$\pm$\,0.011 & 0.84\,$^{+0.16}_{-0.25}$\\
\tableline
\end{tabular}
\end{center}
\end{table}


\begin{thebibliography}{}
\bibitem[2001]{allard01}
  Allard, F., Hauschildt, P.\,H., A., D.\,R., Tamanai, A., \& Schweitzer, 
  A$.$ 2001, \apj, 556, 357
\bibitem[2000]{allen00}
  Allen, W.\,Bb$.$ 2000, Allen's Astrophysical Quantities. Fourth edition, ed$.$ 
  Arthur N$.$ Cox, New York: Springer Verlag, p$.$ 151 
\bibitem[1989]{anders89}
  Anders, E., \& Grevesse, N$.$  1989, GeGoAA, 53, 197
\bibitem[1999]{asiain99}
  Asiain, R., Figueras, F., Torra, J., \& Chen, B$.$ 1999, A\&A, 341, 427
%\bibitem[1999]{asplund99}
%  Asplund, M., Lambert, D.\,L., Kipper, T., Pollacco, D., \&Shetrone, M.\,D$.$ 
%  1999, A\&A, 343, 507
\bibitem[1998]{baraffe98}
  Baraffe, I., Chabrier, G., Allard, F., \& Hauschildt, P.\,H$.$ 1998, \aap, 
  337, 403
\bibitem[2003]{baraffe03}
  Baraffe, I., Chabrier, G., Barman, T.\,S., Allard, F., \& Hauschildt,
  P.\,H$.$ 2003, A\&A, 402, 701
\bibitem[2000]{barbier00}
  Barbier-Brossat, M., \& Figon, P$.$ 2000, A\&AS, 142, 217
\bibitem[1995]{basri95}
  Basri, G., \& Marcy, G.\,W$.$ 1995, \aj, 109, 762
\bibitem[1999]{basri99}
  Basri, G., \& Mart\'\i n, E.\,L$.$ 1999, \aj, 118, 2460
\bibitem[2001]{basri01}
  Basri, G$.$ 2001, \araa, 38, 485 
\bibitem[2003]{burgasser03}
  Burgasser, A.\,J., Kirkpatrick, J.\,D., Reid, I.\,N., 
  Brown, M.\,E., Miskey, C.\,L., \& Gizis, J.\,E$.$ 2003, 
  \apj, 586, 512
\bibitem[1993]{burrows93}
  Burrows, A., Hubbard, W.\,B., Saumon, D., \& Lunine, J.\,I$.$ 1993,
  \apj, 406, 158
\bibitem[1997]{burrows97}
  Burrows, A., et al$.$ 1997, \apj, 491, 856
\bibitem[2000]{burrows00}
  Burrows, A., Marley, M., \& Sharp, C.\,M$.$ 2000, \apj, 531, 438
\bibitem[2002]{burrows02}
  Burrows, A., Ram, R.\,S., Bernath, P., Sharp, C.\,M., \& Milsom, J.\,A$.$
  2002, \apj, 577, 986
\bibitem[2004]{caballero04}
  Caballero, J.\,A., B\'ejar, V.\,J.\,S., Rebolo, R., \& 
  Zapatero Osorio, M.\,R$.$ 2004, A\&A, submitted
\bibitem[2000]{chabrier00}
  Chabrier, G., \& Baraffe, I$.$ 2000, \araa, 38, 337
\bibitem[1997]{chen97}
  Chen, B., Asiain, R., Figueras, F., \& Torra, J$.$ 1997, A\&A, 318, 29
\bibitem[1999]{chereul99}
  Chereul, E., Cr\'ez\'e, M., \& Bienaym\'e, O$.$ 1999, A\&AS, 135, 5
\bibitem[2002]{clarke02}
  Clarke, F.\,J., Tinney, C.\,G., \& Covey, K.\,R$.$ 2002, \mnras, 332, 
  361
\bibitem[2002b]{close02b}
  Close, L.\,M., Potter, D., Brandner, W., Lloyd-Hart, M., Liebert,
  J., Burrows, A., \& Siegler, N$.$ 2002b, \apj, 566, 1095
\bibitem[2002a]{close02a}
  Close, L.\,M., Siegler, N., Potter, D., Brandner, W., \& Liebert, J$.$ 
  2002a, \apj, 567, L53
\bibitem[2003]{close03}
  Close, L.\,M., Siegler, N., Freed, M., \& Biller, B$.$ 2003,
  \apj, 587, 407
\bibitem[2002]{dahn02}
  Dahn, C.\,C., et al$.$ 2002, \aj, 124, 1170
\bibitem[1994]{dantona94}
  D'Antona, F., \& Mazzitelli, I$.$ 1994, \apjs, 90, 467
\bibitem[1997]{dantona97}
  D'Antona, F., \& Mazzitelli, I$.$ 1997, Mem$.$ S$.$ A$.$ It., 68, 807 
\bibitem[2000]{delfosse00}
  Delfosse, X., Forveille, T., S\'egransan, D., Beuzit, J.-L., Udry, S., 
  Perrier, C., \& Mayor, M$.$ 2000, A\&A, 364, 217
\bibitem[2003]{dulick03}
  Dulick, M., Bauschlicher, C.\,W.\,Jr., Burrows, A., Sharp, C.\,M., 
  Ram, R.\,S., \& Bernath, P$.$ 2003, \apj, 594, 651
\bibitem[1969]{eggen69}
  Eggen, O.\,J$.$ 1969, \apj, 155, 701
\bibitem[1992]{eggen92}
  Eggen, O.\,J$.$ 1992, \aj, 104, 1493
\bibitem[1988]{forrest88}
  Forrest, W.\,J., Skrutskie, M.\,F., \& Shure, M$.$ 1988, \apj, 330,
  L119
\bibitem[2002]{gizis02}
  Gizis, J.\,E., Reid, I.\,N., \& Hawley, S.\,L$.$ 2002, \aj, 123, 3356
\bibitem[2003]{gizis03}
  Gizis, J.\,E., et al$.$ 2003, \aj, 125, 3302
\bibitem[2003]{gorlova03}
  Gorlova, N.\,I., Meyer, M.\,R., Rieke, G.\,H., \& Liebert, J$.$ 
  2003, \apj, 593, 1074
\bibitem[1974]{grossman74}
  Grossman, A.\,S., Hays, D., \& Graboske, H.\,C$.$ 1974, \aap, 30, 95
%\bibitem[1994]{goorvitch94}
%  Goorvitch, D$.$ 1994, \apjs, 95, 535
\bibitem[1982]{gurvitz82}
  Gurvitz, L.\,V., Weitz, I.\,V., \& Medvedev, V.\,A$.$ 1982,
  Thermodynamic properties of individual substances. Moscow. Nauka.
\bibitem[1990]{henry90}
  Henry, T.\,J., \& Kirkpatrick, J.\,D$.$ 1990, \apj, 354, L29
\bibitem[1987]{johnson87}
  Johnson, D.\,R.\,H., \& Soderblom, D.\,R$.$ 1987, \aj, 93, 864
\bibitem[2002]{jones02}
  Jones, H.\,R.\,A., Pavlenko, Ya.\,V., Viti, S., \& Tennyson, J$.$ 
  2002, \mnras, 330, 675
\bibitem[2001]{kenworthy01}
  Kenworthy, M., Hofmann, K.-H., Close, L., et al$.$ 2001,
  \apj, 554, L67
\bibitem[2003]{king03}
  King, J.\,R., Villarreal, A.\,R., Soderblom, D.\,R., Gulliver,
  A.\,F., \& Adelman, S.\,J$.$ 2003, \aj, 125, 1980
\bibitem[1963]{kumar63}
  Kumar, H$.$ 1963, \apj, 137, 1121
\bibitem[1999]{kupka99}
  Kupka, F., Piskunov, N., Ryabchikova, T.\,A., Stempels, H.\,C., 
  \& Weiss, W.\,W$.$ 1999, A\&AS, 138, 119
\bibitem[1993]{kurucz93}
  Kurucz, R.\,L$.$ 1993, CDROMs 1-22, Harvard-Smisthonian Observatory
\bibitem[2001b]{lane01b}
  Lane, B.\,F., Boden, A.\,F., \& Kulkarni, S.\,R$.$ 2001b, \apj, 551, L81
\bibitem[2001a]{lane01a}
  Lane, B.\,F., Zapatero Osorio, M.\,R., Britton, M., Mart\'\i n, E.\,L., 
  \& Kulkarni, S$.$ 2001a, \apj, 560, 390, Paper~I
\bibitem[2004]{madsen04}
  Madsen, S., Dravins, D., \& Lindegren, L$.$ 2004, A\&A, in press
\bibitem[1993]{magazzu93}
  Magazz\`u, A., Mart\'\i n, E.\,L., \& Rebolo, R$.$ 1993, \apj, 404, L17
\bibitem[1989]{marcy89}
  Marcy, G.\,W., \& Benitz, K.\,J$.$ 1989, \apj, 344, 441
\bibitem[1987]{marcy87}
  Marcy, G.\,W., Lindsay, V., \& Wilson, K$.$ 1987, \pasp, 99, 490 
\bibitem[1999]{martin99}
  Mart\'\i n, E. L., Brandner, W., \& Basri, G$.$ 1999, Science, 283, 1718
\bibitem[2000]{martin00}
  Mart\'\i n, E.\,L., Koresko, C.\,D., Kulkarni, S.\,R., Lane, B.\,F., 
  \& Wizinowich, P.\,L$.$ 2000, \apj, 529, L37
\bibitem[2003]{martin03}
  Mart\'\i n, E.\,L., Barrado y Navascu\'es, D., Baraffe, I., 
  Bouy, H., \& Dahm, S$.$ 2003, \apj, 594, 525 
\bibitem[2004]{mccaughrean04}
  McCaughrean, M.\,J., et al$.$ 2004, A\&A, 413, 1029
\bibitem[2004]{mcgovern04}
  McGovern, M.\,R., Kirkpatrick, J.\,D., McLean, I.\,S., Burgasser, 
  A.\,J., Prato, L., \& Lowrance, P.\,J$.$ 2004, \apj, 600, 1020
\bibitem[1998]{mclean98}
  McLean, I.\,S., et al$.$ 1998, SPIE, 3354, 566
\bibitem[2004]{mohanty04}
  Mohanty, S., Basri, G., Jayawardhana, R., Allard, F., 
  Hauschildt, P., \& Ardila, D$.$ 2004, \apj, in press
\bibitem[2001]{montes01}
  Montes, D., L\'opez-Santiago, J., G\'alvez, M.\,C., 
  Fern\'andez-Figueroa, M.\,J., de Castro, E., \& Cornide, M$.$ 2001, 
  \mnras, 328, 45
\bibitem[1995]{nakajima95}
  Nakajima, T., Oppenheimer, B.\,R.,  Kulkarni, S.\,R., Golimowski, 
  D.\,A., Matthews, K., \& Durrance, S.\,T$.$ 1995, \nat, 378, 463
\bibitem[2002]{nidever02}
  Nidever, D.\,L., Marcy, G.\,W., Butler, R.\,P., Fischer, D.\,A., 
  \& Vogt, S.\,S$.$ 2002, \apjs, 141, 503
\bibitem[1995]{orlov95}
  Orlov, V.\,V., Panchenko, I.\,E., Rastorguev, A.\,S., \& Yatsevich, 
  A.\,V$.$ 1995, AZh, 72, 495 (English transl$.$ Astron$.$ Rep$.$, 
  39, 437
\bibitem[1997]{partrige97}
  Partrige, H., \& Schwenke, D.\,W$.$ 1997, J$.$ Chem$.$ Phys., 
  106, 4618
\bibitem[2000]{pav00}
  Pavlenko, Ya.\,V$.$ 1997, Astron$.$ Rep, 44, 219
\bibitem[2002]{pav02}
  Pavlenko, Ya.\,V$.$ 2002, Astron$.$ Rep, 46, 567
\bibitem[2002]{pavjones02}
  Pavlenko, Ya.\,V., \& Jones, H.\,R.\,A$.$ 2002, A\&A, 396, 967
\bibitem[2004]{pav04}
  Pavlenko, Ya.\,V., Geballe, T.\,R., Evans, A., Smalley, B., 
  Eyres, S.\,P.\,S., Tyne, V.\,H., \& Yakovina, L.\,A$.$ 2004, A\&A, 
  in press
\bibitem[1997]{perryman97}
  Perryman, M.\,A.\,C., et al$.$ 1997, A\&A, 323, L49
\bibitem[1998]{plez98}
  Plez, B$.$ 1998, A\&A, 337, 495
\bibitem[2002]{potter02} 
  Potter, D., Mart\'\i n, E.\,L., Cushing, M.\,C., Baudoz, P., 
  Brandner, W., Guyon, O., \& Neuh\"auser, R$.$ 2002, \apj, 567, L133
\bibitem[1995]{rebolo95}
  Rebolo, R., Zapatero Osorio, M.\,R., \& Mart\'\i n, E.\,L$.$ 1995, 
  \nat, 377, 129
\bibitem[2002]{reid02}
  Reid, I.\,N., Kirkpatrick, J.\,D., Liebert, J., Gizis, J.\,E., Dahn, C.\,Bb., 
  \& Monet, D.\,G$.$ 2002, \aj, 124, 519
\bibitem[2003]{ribas03}
  Ribas, I$.$ 2003, A\&A, 398, 239
\bibitem[2003]{segransan03}
  S\'egransan, D., Kervella, P., Forveille, T., \& Queloz, D$.$ 2003,
  A\&A, 397, L5
\bibitem[1993]{soderblom_stauffer93}
  Soderblom, D.\,R., Stauffer, J.\,R., Hudon, J.\,D., \& Jones, B.\,F$.$
  1993, \apjs, 85, 315
\bibitem[1993]{soderblom93}
  Soderblom, D.\,R., \& Mayor, M$.$ 1993, \aj, 105, 226
\bibitem[1998]{tinney98}
  Tinney, C.\,G., \& Reid, I.\,N$.$ 1998, \mnras, 301, 1031
\bibitem[1986]{tomkin86}
  Tomkin, J., \& Popper, D.\,M$.$ 1986, \aj, 91, 1428
\bibitem[1973]{tsuji73}
  Tsuji, T$.$ 1973, A\&A, 23, 411
\bibitem[2002]{tsuji02}
  Tsuji, T$.$ 2002, \apj, 575, 264
\bibitem[1955]{unsold55}
  Uns\"old, A$.$ 1955, Physik der Sternatmospheren, 2nd ed., Springer.
  Berlin
\bibitem[1988]{wizinowich88}
  Wizinowich, P.\,L., et al$.$ 1988, Proc$.$ SPIE, 3353, 568
\bibitem[2003]{osorio03}
  Zapatero Osorio, M.\,R., Caballero, J.\,A., B\'ejar, V.\,J.\,S., 
  \& Rebolo, R$.$ 2003, A\&A, 408, 663
\end{thebibliography}
\end{document}